\documentclass[12pt]{article}
\pdfoutput=1
\usepackage[utf8]{inputenc}
\usepackage{cite}
\usepackage{amsmath,amssymb,amsbsy,amstext,amsthm,simplewick,amsfonts}
\usepackage{graphicx}
\usepackage{wrapfig}
\usepackage{upgreek}
\usepackage{bm} %--> bold math...
\usepackage{framed}
\usepackage{bbm}
\usepackage{textcomp}
\usepackage{tikz}
\usepackage{pifont}
\usetikzlibrary{matrix,shapes,fit,tikzmark,calc}
\usepackage{adjustbox}
\usepackage{makecell}
\usepackage{tcolorbox}
\usepackage{physics}
\usepackage{empheq}
\usepackage[normalem]{ulem}
\usepackage{enumitem}
\usepackage{braket} 
\usepackage{array}
\usepackage{dsfont}
\usepackage{ulem}
\usepackage{titlesec}
\numberwithin{equation}{section}

\usepackage{caption}
\usepackage{subcaption}
\usepackage[framemethod=default]{mdframed}
\usepackage{colortbl}

\usepackage{hyperref}
\hypersetup{
    colorlinks=true,
    linkcolor=red,
    citecolor=blue,
} 

\makeatletter
\newlength{\apb@width}
\newcommand{\autoparbox}[2][c]{\settowidth{\apb@width}{#2}\parbox[#1]{\apb@width}{#2}}

\makeatother


\def\D{\mathrm d}
\def\vec{\bm}
\def\e{\mathrm e}

\allowdisplaybreaks[1]
\begin{document}

%\newgeometry{top=2cm, bottom=2cm, left=2.9cm, right=2.9cm}

\begin{titlepage}
\setcounter{page}{1} \baselineskip=15.5pt

\thispagestyle{empty}

\renewcommand*{\thefootnote}{\fnsymbol{footnote}}

\begin{center}

{\fontsize
{20}{20} \bf Scalar-Induced Gravitational Waves

\medskip

from Ghost Inflation and Parity Violation\;} \\
\end{center}

\vskip 18pt
\begin{center}
\noindent
{\fontsize{12}{18}\selectfont Sebastian Garcia-Saenz\footnote{ sgarciasaenz@sustech.edu.cn}, Yizhou Lu\footnote{ luyz@sustech.edu.cn},
Zhiming Shuai\footnote{12132939@mail.sustech.edu.cn}
}
\end{center}

\begin{center}
\vskip 8pt
\textit{ Department of Physics, Southern University of Science and Technology, Shenzhen 518055, China} \\ 
\end{center}

%=========================================

\vspace{1.4cm}

\noindent  We calculate the scalar-induced gravitational wave energy density in the theory of Ghost Inflation, assuming scale invariance and taking into account both the power spectrum- and trispectrum-induced contributions. For the latter we consider the leading cubic and quartic couplings of the comoving curvature perturbation in addition to two parity-violating quartic operators. In the parity-even case, we find the relative importance of the trispectrum-induced signal to be suppressed by the requirement of perturbativity, strengthening a no-go theorem recently put forth. The parity-odd signal, even though also bound to be small, is non-degenerate with the Gaussian contribution and may in principle be comparable to the parity-even non-Gaussian part, thus potentially serving as a probe of the Ghost Inflation scenario and of parity violating physics during inflation.

%=========================================

\end{titlepage}

\setcounter{tocdepth}{2}
{
\hypersetup{linkcolor=black}
\vspace{10pt}
\noindent\rule{\textwidth}{1pt}
\tableofcontents
\noindent\rule{\textwidth}{1pt}
\vspace{10pt}
}

\renewcommand*{\thefootnote}{\arabic{footnote}}
\setcounter{footnote}{0} 
\newpage

\section{Introduction}

Cosmic inflation has the potential of providing a wealth of information on physics at energy scales not achievable by artificial particle colliders. One such effect is the violation of parity invariance by high-energy interactions which, being a symmetry, is of prime importance in the effort to constrain fundamental theories with experimental data. In fact hints of parity violating physics at play in the early universe have already been discovered in the statistics of the large-scale structure \cite{Philcox:2022hkh,Hou:2022wfj}, and so it seems crucial to understand whether this breaking of parity could have a primordial origin in the context of inflation.

The most minimal theory for the physics of fluctuations during inflation is described by a scalar mode, physically encoding the clock that will dictate the end of inflation, and two tensor modes corresponding to the perturbations of the gravitational field \cite{Cheung:2007st,Piazza:2013coa}. 
 The former is most conveniently parametrized by the comoving curvature perturbation $\zeta$. Signatures of parity violation will in principle be manifest in correlation functions of these fluctuations and one would expect the most important effects to arise in the non-Gaussian part of the 4-point function of $\zeta$ and in the 2-point function of the graviton.\footnote{Due to homogeneity and isotropy, the scalar 2-point function cannot break parity because it is a function of only the magnitude of the momentum. For the same reason, the momenta in the scalar 3-point function must form a triangle, and a triangle is unchanged by a parity transformation. Thus the leading effect in the scalar sector appears in the 4-point function.}

The experimental measurement of these correlators is nevertheless challenging in either case. At least on scales probed by the cosmic microwave background (CMB), the probability distribution of $\zeta$ is inferred to be extremely Gaussian, making it unlikely to achieve a detection of parity violation in the foreseeable future \cite{Planck:2019kim,Planck:2015zfm,Renaux-Petel:2015bja}, while in the case of primordial gravitational waves (GWs) even a Gaussian signal is constrained to be minute \cite{Planck:2018vyg,BICEP:2021xfz}. However both $\langle\zeta^4\rangle$ and $\langle h^2\rangle$ (where $h$ denotes, schematically for now, the tensor fluctuation) are essentially unconstrained on scales much smaller than those probed in the CMB, and especially exciting is the possibility of observing a primordial GW signal in forthcoming experiments \cite{TianQin:2015yph,Bartolo:2016ami,Hu:2017mde,NANOGrav:2020bcs,LISACosmologyWorkingGroup:2022jok}.

Focusing therefore on the tensor 2-point function as a particularly well-motivated probe of parity breaking physics during inflation \cite{Lue:1998mq,Contaldi:2008yz,Maldacena:2011nz,Bartolo:2018elp,Orlando:2020oko,Orlando:2022rih}, we can envisage two distinct dynamical sources of parity violation, namely in the gravitational sector or in the scalar sector.\footnote{More possibilities are of course available in theories with multiple fields. One such class of models that predicts a breaking of parity invariance is axion inflation \cite{Freese:1990rb,Adshead:2012kp}.} Parity-odd tensor couplings have received considerable recent interest in connection to this question \cite{Takahashi:2009wc,Soda:2011am,Creminelli:2014wna,Bartolo:2017szm,Qiao:2019hkz,Bordin:2020eui,Bartolo:2020gsh,Cabass:2021fnw,Cai:2021uup,Li:2022mti,Li:2022vtn,Zhang:2022xmm,Feng:2023veu,Creque-Sarbinowski:2023wmb,Rao:2023doc,Zhu:2023wci,Stefanyszyn:2023qov}. Yet models of gravity that modify general relativity, including in particular those that feature parity violation, face several issues related to their stability --- in the case of higher-curvature theories --- in addition to stringent experimental tests, so that further studies are necessary in order to assess their viability. Here instead we focus on the possibility that the breaking of parity arises exclusively from the scalar sector of the model, with the gravitational interactions being the standard ones of Einstein's theory.

Scalar fluctuations contribute to the tensor 2-point function at second order in perturbation theory through mixings of the form $h_{ij}\partial_i\zeta\partial_j\zeta$. The resulting signal, known as scalar-induced GWs (SIGW), has been the subject of intense scrutiny in the past years as several mechanisms to enhance scalar perturbations on small scales have been discovered \cite{Saito:2008em,Garcia-Bellido:2017mdw,Garcia-Saenz:2018ifx,Cai:2018tuh,Garcia-Saenz:2018vqf,Cai:2019jah,Lin:2020goi,Fumagalli:2020adf,Yi:2020cut,Fumagalli:2020nvq,Gao:2020tsa,Gao:2021dfi,Fumagalli:2021cel,Witkowski:2021raz,Fumagalli:2021mpc}; see also \cite{Domenech:2021ztg} for a review and further references.\footnote{Although outside the scope of this paper, it is worth mentioning that SIGWs have the notorious issue of gauge dependence \cite{Hwang:2017oxa,DeLuca:2019ufz,Inomata:2019yww}. See also \cite{Domenech:2021ztg} for a summary and further references.} In general, the SIGW power spectrum receives two contributions, schematically $\langle h^2\rangle\sim\langle\zeta^4\rangle_{\rm d}+\langle\zeta^4\rangle_{\rm c}$, where $\langle\zeta^4\rangle_{\rm d}$ denotes the ``disconnected'' or Gaussian part of the scalar 4-point correlator and $\langle\zeta^4\rangle_{\rm c}$ denotes the ``connected'' or non-Gaussian component, i.e.\ the scalar trispectrum.
See \cite{Cai:2018dig,Unal:2018yaa,Adshead:2021hnm,Garcia-Saenz:2022tzu,Li:2023qua} for studies on the impact of non-Gaussianities on SIGWs.
While one typically expects the connected contribution to be subleading relative to the disconnected one (more on this below), the two have in general a different scale dependence and are therefore in principle distinguishable. The trispectrum-induced component is of course especially interesting as it encodes direct information about the interactions of the curvature perturbation during inflation and, more to the point of this paper, it is only this component that can a priori produce a breaking of parity invariance.

Trispectrum-induced GWs have been the subject of some interesting recent studies \cite{Unal:2018yaa,Atal:2021jyo,Adshead:2021hnm}, although these works were restricted to local-type non-Gaussianities. However it was later shown in \cite{Garcia-Saenz:2022tzu} that the requirement of perturbativity necessarily bounds the relative importance of the local trispectrum-induced GW signal to be small. More in detail, the ratio $\Omega_{\rm GW,c}/\Omega_{\rm GW,d}$ of the connected and disconnected SIGW contributions is of the order of $P_{\zeta}^{\rm(1-loop)}/P_{\zeta}^{\rm(tree)}$, i.e.\ the ratio of the 1-loop scalar power spectrum to its tree level result, which must be parametrically smaller than one for the theory to be under perturbative control. Moreover, this result was also shown to apply to the regular trispectrum shapes that result from derivative interactions in single-field slow-roll inflation as well as models with reduced speed of sound. It should be remarked, however, that the results of \cite{Garcia-Saenz:2022tzu} were derived under the assumption of scale invariance.

The no-go theorem of \cite{Garcia-Saenz:2022tzu} may a priori still be evaded by other trispectrum shapes. Here we investigate two classes of trispectra that arise in the theory of Ghost Inflation \cite{Arkani-Hamed:2003pdi,Arkani-Hamed:2003juy,Arkani-Hamed:2005teg}. Our motivation to investigate this scenario is three-fold. First, we would like to assess the robustness of the above no-go result by studying a model of inflation which differs from the standard single-field description, but with which it nevertheless shares some similarities, namely derivative interactions and equilateral-type non-Gaussianities. Second, Ghost Inflation has been identified as viable mechanism to enhance the scalar power spectrum in a manner consistent with perturbative unitarity \cite{Ballesteros:2021fsp}, thus serving as a well-motivated arena to investigate not only SIGWs but also primordial black hole formation \cite{Hawking:1971ei,Carr:1974nx,Sasaki:2018dmp}. Third, the breaking of parity in the scalar trispectrum turns out to be impossible in models with linear dispersion relation and Bunch-Davies initial conditions, assuming scale invariance \cite{Liu:2019fag,Cabass:2022rhr}.\footnote{This statement holds for the tree-level trispectrum but is generically evaded at one-loop order \cite{Lee:2023jby}.}

The last point is crucial since, as we already mentioned, it is only a parity-odd trispectrum that can in principle lead to a violation of parity in the SIGW power spectrum, again assuming the gravitational sector is not modified. Ghost Inflation evades the no-go result of \cite{Cabass:2022rhr} precisely because the dispersion relation is non-linear, $\omega\propto k^2$. Although Ghost Inflation predicts scale-invariant correlation functions (see however \cite{Senatore:2004rj}), what we actually have in mind is a scenario in which the dispersion relation of the adiabatic mode changes from linear to non-linear on small scales. In this way, it is possible to have an enhancement of the scalar power spectrum on small scales relative to the one constrained by CMB observations on large scales \cite{Ballesteros:2021fsp}. In this set-up, the physics of Ghost Inflation is therefore understood to apply only on small enough scales.\footnote{Power-spectrum induced GWs in a model with modified dispersion relation have been studied in \cite{Qiu:2022klm}.}

In this paper we calculate the SIGW power spectrum for four different scalar trispectra in the theory of Ghost Inflation and with the assumption of exact scale invariance.
Two of the trispectra are parity-even and correspond to the most relevant self-interactions of the curvature perturbation, namely a cubic operator and a quartic operator that contribute to the connected 4-point function through scalar-exchange and contact diagrams, respectively \cite{Huang:2010ab,Izumi:2010wm}.\footnote{See also \cite{Senatore:2010jy,Ashoorioon:2021srt,Ashoorioon:2011eg} for other studies of non-Gaussianities in models with non-linear dispersion relation.} The second set of trispectra is given by two parity-odd quartic operators which correspond to the leading sources of parity violation in the scalar sector \cite{Cabass:2022oap,Cabass:2022rhr}. Our results are compared with the Gaussian contribution to the SIGW spectrum upon taking into account the condition of perturbative control. Let us summarize our main results:
\begin{itemize}
    \item For the parity-even trispectra, we find a stringent bound on the relative importance of the connected SIGW spectrum allowed by perturbativity, thereby confirming and generalizing the no-go theorem of \cite{Garcia-Saenz:2022tzu} to encompass scenarios with modified dispersion relation. The bounds are comparable or even stronger than in slow-roll inflation or in models with reduced speed of sound.
    
    \item For the parity-odd trispectra, we find a very small result, of $\mathcal{O}(10^{-3})$ times what naive dimensional analysis would predict. The degree of chirality is bounded by a number parametrically smaller than $\mathcal{O}(10^{-1})$ from the condition of perturbativity. We argue that, optimistically, this could be of the same order as the parity-even trispectrum-induced contributions.
\end{itemize}

In Sec.\ \ref{sec:SIGW} we provide a brief review of the computational aspects of SIGWs. We pay special attention to identify the conditions under which the trispectrum-induced signal may be non-vanishing. We explain that (i) only trispectra that are \textit{even} functions of the scalar product $(\bm q_1\cdot \bm q_2)$ provide a net non-zero contribution to the total GW power spectrum (here $\bm q_{1,2}$ are the internal momenta to be integrated over in the Green function convolution); (ii) only trispectra that are \textit{odd} functions of $(\bm k\cdot\bm q_1\times \bm q_2)(\bm q_1\cdot \bm q_2)$ can produce a parity-violating GW signal, i.e.\ with different power spectra for left- and right-handed GW polarizations. (These statements assume that the trispectrum is a polynomial in the scalar products $(\bm q_1\cdot \bm q_2)$ and $(\bm k\cdot\bm q_1\times \bm q_2)$, however analogous criteria may also be inferred in more general cases.) An interesting implication is that a trispectrum function that depends on $(\bm q_1\cdot \bm q_2)$ \textit{only} through an odd function of $(\bm k\cdot\bm q_1\times \bm q_2)(\bm q_1\cdot \bm q_2)$ will yield zero net power in GWs and is therefore only detectable in experiments that can measure individual chiral polarizations. Indeed, we also explain that linearly polarized GWs are not sensitive to parity violation in the scalar trispectrum. These observations are precisely relevant in the case of Ghost Inflation, as we will see.

In Sec.\ \ref{sec:GI}, after a short review of Ghost Inflation and the 4-point correlation functions that result from this scenario, we present our calculation of the GW power spectrum induced by the scalar trispectra mentioned previously and derive our main results. We conclude in Sec.\ \ref{sec:conc} with some final comments. We collect in the Appendices some additional information and results related to our conventions and numerical calculations.

%%%%%%%%%%%%%%%%%%%%%%%%%%%%%%%%%%%%%
%%%%%%%%%%%%%%%%%%%%%%%%%%%%%%%%%%%%%

\section{Scalar-induced GWs}\label{sec:SIGW}

This section presents a brief overview of the theory of scalar-induced GWs, followed by a detailed analysis of the master integral that defines the scalar trispectrum contribution for different GW polarizations. Based on the structure of the master integral and the symmetries of the theory, we identify necessary conditions for the trispectrum to yield a non-vanishing signal both in the total power as well as in individual polarization channels. In particular, we explain how a parity-odd trispectrum leads to different SIGW spectra for chiral polarizations.

\subsection{Elements of SIGWs}

We follow the conventions of \cite{Domenech:2021ztg} to which we refer the reader for more details. We define the tensor perturbation $h_{ij}(\bm x,\eta)$ by
\begin{align}\label{h_def}
    \D s^2=a(\eta)^2[-\D\eta^2+(\delta_{ij}+h_{ij}(\vec x,\eta))\D x^i\D x^j],
\end{align}
in terms of conformal time $\eta$. Expanding the tensor perturbation in Fourier modes and polarizations, we have
\begin{equation}
    h_{ij}(\vec x,\eta)=\sum_{\lambda=+,\times}\int\frac{\D^3\vec k}{(2\pi)^3}\e_{ij}^\lambda(\hat{\vec k})h_{\bm k}^\lambda(\eta)e^{i\vec k\cdot x},
\end{equation}
where the index $\lambda$ labels the GW polarization. For $+$ and $\times$ polarizations we define the tensors as
\begin{align}
    \e^+_{ij}(\hat{\bm k})=&\, \frac{1}{\sqrt2}[\e_i(\hat{\bm k})\e_j(\hat{\bm k})-\bar{\e}_i(\hat{\bm k})\bar{\e}_j(\hat{\bm k})],\\
    \e^\times_{ij}(\hat{\bm k})=&\, \frac{1}{\sqrt2}[\e_i(\hat{\bm k})\bar{\e}_j(\hat{\bm k})+\bar{\e}_i(\hat{\bm k}){\e}_j(\hat{\bm k})],
\end{align}
where $\e_i(\hat{\vec k})$ and $\bar{\e}_i(\hat{\vec k})$ are orthonormal vectors, orthogonal to $\hat{\vec k}$ and to each other, that is $\e_i\e_i=\bar{\e}_i\bar{\e}_i=1$ and $\e_i\bar{\e}_i=0$. Some useful properties of these tensors are
\begin{align}
    k_i\e_{ij}^{+,\times}&=0,\\
    \e_{ij}^+\e_{ij}^+&=\e_{ij}^\times\e_{ij}^{\times}=1.
\end{align}

The tensor power spectrum $P_{\lambda}$ is defined by
\begin{equation}\label{eq:tensor power spectrum def}
    \braket{h_{\bm k}^\lambda(\eta)h_{\bm k'}^{\lambda'}(\eta)}=(2\pi)^3
    \delta(\bm k+\bm k')\delta^{\lambda\lambda'} P_\lambda(k,\eta).
\end{equation}
The energy density per logarithmic wavelength of GWs is given by \cite{Isaacson:1968hbi,Isaacson:1968zza,Maggiore:1999vm,Kohri:2018awv}
\begin{equation}
    \Omega_{\rm GW}(k,\eta)=\frac{1}{12}\left(\frac{k}{\mathcal H}\right)^2\sum_{s=+,\times}\overline{\mathcal P_\lambda(k,\eta)},
\end{equation}
where $\mathcal P_\lambda=\frac{k^3}{2\pi^2}P_\lambda$ is the dimensionless power spectrum, the overline means an average over many oscillations of the GWs and
$\mathcal H=aH$ is the conformal Hubble parameter.

In Einstein gravity, the equation of motion (EOM) for the tensor modes to second order in perturbation theory reads
\begin{equation}\label{eq:tensor eom}
    h_{\bm k}^{\lambda\prime\prime}(\eta)+2\mathcal{H}h_{\bm k}^{\lambda\prime}(\eta)+k^2h_{\bm k}^{\lambda}(\eta)=2\mathcal S_{\bm k}^{\lambda}(\eta),
\end{equation}
where the primes indicate derivatives with respect to conformal time $\eta$, and $\mathcal{S}_{\vec k}^\lambda$ is the source of quadratic order in $\zeta$ projected onto the $\lambda$ polarization. In Newtonian gauge it is given by \cite{Ananda:2006af,Baumann:2007zm,Kohri:2018awv,Espinosa:2018eve}\footnote{See \cite{Lu:2020diy} for the general expression in an arbitrary gauge.}
\begin{equation}
    \mathcal S_{\bm{k}}=\int \frac{\mathrm{d}^3 q}{(2 \pi)^{3}} \e_{i j}(\hat{\bm{k}}) q_i q_j\left(2 \Phi_{\bm{q}} \Phi_{\bm{k}-\bm{q}}+\frac{4}{3(1+w)}\left(\mathcal{H}^{-1} \Phi_{\bm{q}}^{\prime}+\Phi_{\bm{q}}\right)\left(\mathcal{H}^{-1} \Phi_{\bm{k}-\bm{q}}^{\prime}+\Phi_{\bm{k}-\bm{q}}\right)\right),
\end{equation}
where the Bardeen potential $\Phi$ is related to the comoving curvature perturbation $\zeta$ on super-Hubble scales as $\Phi=\frac{3+3w}{5+3w} \zeta$ with $w$ the equation-of-state parameter. We will focus on the radiation-dominated era so that $w=1/3$ and $\Phi=\frac{2}{3}\zeta$.

Solving the EOM with the Green function method and neglecting a primordial contribution, the tensor power spectrum is found to be \cite{Ananda:2006af,Baumann:2007zm,Inomata:2016rbd,Kohri:2018awv}
\begin{equation} \label{h_powerspec}
    \begin{aligned}
P_\lambda(\eta, k)= &\, \frac{4}{k^4}\left(\frac{2}{3}\right)^4 \int \frac{\mathrm{d}^3 \boldsymbol{q}_1}{(2 \pi)^3} \frac{\mathrm{d}^3 \boldsymbol{q}_2}{(2 \pi)^3} Q_\lambda\left(\boldsymbol{k}, \boldsymbol{q}_1\right) Q_\lambda\left(-\boldsymbol{k}, \boldsymbol{q}_2\right) \\
& \times I\left(\frac{\left|\boldsymbol{k}-\boldsymbol{q}_1\right|}{k}, \frac{q_1}{k}, k \eta\right) I\left(\frac{\left|\boldsymbol{k}-\boldsymbol{q}_2\right|}{k}, \frac{q_2}{k}, k \eta\right)\left\langle\zeta_{\boldsymbol{q}_1} \zeta_{\boldsymbol{k}-\boldsymbol{q}_1} \zeta_{-\boldsymbol{q}_2} \zeta_{-\boldsymbol{k}+\boldsymbol{q}_2}\right\rangle^{\prime},
\end{aligned}
\end{equation}
where
\begin{equation}\label{eq:Q def}
    Q_\lambda(\vec k,\vec q)\equiv \e_{ij}^\lambda(\hat{\vec k})q_iq_j.
\end{equation}
The function $I(u,v,x\equiv k\eta)$ contains the post-inflation evolution information about the scalar source and may be found e.g.\ in \cite{Kohri:2018awv,Adshead:2021hnm}, and
$\braket{\cdots}'$ implies that $(2\pi)^3\delta(\sum \vec k_a)$ in the correlator has been removed. It represents the primordial 4-point function of $\zeta$ generated during inflation.

In general, the 4-point correlation function can be decomposed into connected and disconnected contributions,
\begin{equation}
    \braket{\zeta_{\bm k_1}\zeta_{\bm k_2}\zeta_{\bm k_3}\zeta_{\bm k_4}}'=\braket{\zeta_{\bm k_1}\zeta_{\bm k_2}\zeta_{\bm k_3}\zeta_{\bm k_4}}'_{\rm c}+\braket{\zeta_{\bm k_1}\zeta_{\bm k_2}\zeta_{\bm k_3}\zeta_{\bm k_4}}'_{\rm d},
\end{equation}
where the disconnected part is by definition determined only by the 2-point function, i.e.\ the primordial scalar power spectrum. The connected part instead corresponds to an intrinsic non-Gaussianity in the probability distribution of $\zeta$. Explicitly we have
\begin{align}
    \braket{\zeta_{\bm q_1}\zeta_{\bm k-\bm q_1}\zeta_{-\vec q_2}\zeta_{-\vec k+\bm q_2}}'_{\rm d}&=(2\pi)^3[\delta(\bm q_1-\bm q_2)P_\zeta(q_1)P_\zeta(|\bm k-\bm q_1|)+\delta(\bm q_1+\bm q_2-\bm k)P_\zeta(q_1)P_{\zeta}(q_2)],\label{disconnected}\\
    \braket{\zeta_{\bm q_1}\zeta_{\bm k-\bm q_1}\zeta_{-\vec q_2}\zeta_{-\vec k+\bm q_2}}'_{\rm c}&=T_\zeta(\bm q_1,\bm k-\bm q_1,-\bm q_2,-\bm k+\bm q_2),\label{connected}
\end{align}
where $T_\zeta(\bm k_1,\bm k_2,\bm k_3,\bm k_4)$ is the trispectrum function. It is convenient to define a dimensionless trispectrum as
\begin{equation}
    \mathcal T_\zeta(\bm k_1,\bm k_2,\bm k_3,\bm k_4)=\frac{(k_1k_2k_3k_4)^{9/4}}{(2\pi)^6}T_\zeta(\bm k_1,\bm k_2,\bm k_3,\bm k_4).
\end{equation}
After some manipulations in the expression \eqref{h_powerspec} for the tensor power spectrum we obtain \cite{Garcia-Saenz:2022tzu}
\begin{align}
\overline{\mathcal{P}_{h,\mathrm{d}}}= & \left(\frac{\mathcal H}{k}\right)^2 \int_0^{\infty} \mathrm{d} v \int_{|1-v|}^{1+v} \mathrm{~d} u \, \mathcal{K}_{\mathrm{d}}(u, v) \mathcal{P}_\zeta(k u) \mathcal{P}_\zeta(k v), \label{Ph_disconnected}\\
\overline{\mathcal{P}_{h, \mathrm{c}}}= & \left(\frac{\mathcal H}{k}\right)^2 \int_0^{\infty} \mathrm{d} v_1 \int_{\left|1-v_1\right|}^{1+v_1} \mathrm{~d} u_1 \int_0^{\infty} \mathrm{d} v_2 \int_{\left|1-v_2\right|}^{1+v_2} \mathrm{~d} u_2 \int_0^{2 \pi} \mathrm{d} \psi \,\mathcal{K}_{\mathrm{c}}\left(u_1, v_1, u_2, v_2\right)\label{Ph_connected} \\
& \times \frac{\cos (2 \psi)}{\pi} \mathcal{T}_\zeta\left(u_1, v_1, u_2, v_2, \psi\right)\notag,
\end{align}
which we refer to as the ``master integrals'' that determine respectively the disconnected and connected contributions to the dimensionless power spectrum of GWs. Note that a sum over polarizations has been done, so that these expressions give the total spectrum. The dimensionless integration variables are given by
\begin{equation}
    u_i\equiv \frac{|\bm k-\bm q_i|}{k},\quad v_i\equiv \frac{q_i}{k},
\end{equation}
and we introduced the integration kernels
\begin{align}
\mathcal{K}_{\mathrm{d}}(u, v)= & \left[\frac{4 v^2-\left(1+v^2-u^2\right)^2}{{4} u v}\right]^2 I_A^2(u, v)\left[I_B^2(u, v)+I_C^2(u, v)\right], \\
\mathcal{K}_{\mathrm{c}}\left(u_1, v_1, u_2, v_2\right)= &\, \frac{1}{{16}\left(u_1 v_1 u_2 v_2\right)^{5 / 4}}\left[4 v_1^2-\left(1+v_1^2-u_1^2\right)^2\right]\left[4 v_2^2-\left(1+v_2^2-u_2^2\right)^2\right] \\
& \times I_A\left(u_1, v_1\right) I_A\left(u_2, v_2\right)\left[I_B\left(u_1, v_1\right) I_B\left(u_2, v_2\right)+I_C\left(u_1, v_1\right) I_C\left(u_2, v_2\right)\right],\notag
\end{align}
with $I_{A,B,C}$ being
\begin{align}
    I_A(u,v)&=\frac{3(u^2+v^2-3)}{4u^3v^3},\\
    I_B(u,v)&=-4uv+(u^2+v^2-3)\log\left|
    \frac{3-(u+v)^2}{3-(u-v)^2}
    \right|,\\
    I_C(u,v)&=\pi(u^2+v^2-3)\Theta(u+v-\sqrt3).
\end{align}

\subsection{The polarizations of SIGWs} \label{polar_SIGWs}

The above master integrals correspond to the total power in GWs after summing over polarizations. However in principle an experiment might be able to detect individual polarization signals. This is true in particular for chiral polarizations which give the most direct way to quantify the breaking of parity in GWs.

So let us rewind back to Eq.\ \eqref{h_def} and now consider the expansion of the tensor perturbation in chiral polarizations,
\begin{equation}
    h_{ij}(\vec x,\eta)=\sum_{\lambda=R,L}\int\frac{\D^3\vec k}{(2\pi)^3}\e_{ij}^\lambda(\hat{\vec k})h_{\bm k}^\lambda(\eta)e^{i\vec k\cdot x},
\end{equation}
where the right- and left-handed polarization tensors are defined by
\begin{align}
    \e_{ij}^{R}(\hat{\vec k})&\equiv\frac{1}{\sqrt2}[\e_{ij}^+(\hat{\vec k})+i\e_{ij}^\times(\hat{\vec k})],\\
    \e_{ij}^L(\hat{\vec k})&\equiv \frac{1}{\sqrt2}[\e_{ij}^+(\hat{\vec k})-i\e_{ij}^\times(\hat{\vec k})].
\end{align}
Some useful properties of these tensors are
\begin{align}
    k_i\e_{ij}^{R,L}&=0,\\
    \e_{ij}^{R}\e_{ij}^{R}&=\e_{ij}^{L}\e_{ij}^{L}=0,\\
    \e_{ij}^R(\hat{\vec k})\e_{ij}^L(\hat{\vec k})&=1,\\
    \e_{ij}^R(\hat{\vec k})&=\e_{ij}^{L*}(\hat{\vec k})=\e_{ij}^L(-\hat{\vec k}),\\
    \e_{ij}^L(\hat{\vec k})&=\e_{ij}^{R*}(\hat{\vec k})=\e_{ij}^R(-\hat{\vec k}).
\end{align}

The power spectrum for a given chiral mode is again defined as in \eqref{eq:tensor power spectrum def} with $\lambda=R,L$, while the mode functions $h^{R,L}_{\bm k}$ satisfy the EOM \eqref{eq:tensor eom} with the source function projected onto the chiral basis. It follows that the power spectrum for $h^{R,L}_{\bm k}$ is also given by \eqref{h_powerspec}.

It will be useful for what follows to have explicit expressions for the function $Q_\lambda(\vec k,\vec q)$ introduced in \eqref{eq:Q def}. To this end it is convenient to employ spherical coordinates defined relative to $\vec k=k\hat{\vec z}$,
so that the components of $\vec q$ are given by $\vec q=q[\sin\theta\cos\phi,\sin\theta\cos\phi,\cos\theta]$ in this frame. We then obtain
\begin{equation}
\begin{aligned}
    Q_\lambda(\vec k,\vec q)&=\frac{1}{\sqrt2}q^2\sin^2\theta\times
    \begin{cases}
        \cos2\phi\quad \lambda=+\\\sin2\phi\quad \lambda=\times
    \end{cases} \,,\\
    Q_\lambda(\vec k,\vec q)&=\frac{1}{2}q^2\sin^2\theta\times\begin{cases}
        e^{2i\phi}\quad \lambda=R\\
        e^{-2i\phi}\quad \lambda=L
    \end{cases}.
\end{aligned}
\end{equation}
$Q_\lambda(\bm k,\bm q)$ verifies the following symmetry properties:
\begin{align}
    Q_{\lambda}(\bm k,\bm q+c\bm k)&=Q_\lambda(\bm k,\bm q),\quad \lambda=+,\times,R,L,\\
    Q_{+,\times}(\bm k,\bm q)&=Q_{+,\times}(-\bm k,\bm q)=Q_{+,\times}(\bm k,-\bm q),\\
    Q_{R,L}(\bm k,\bm q)&=Q_{L,R}(-\bm k,\bm q)=Q_{L,R}^*(\bm k,\bm q).
\end{align}
We see that the polar angle dependence is the same for different polarizations, and this is the fact that GWs are transverse. 
Therefore any asymmetry between polarizations can only emerge from azimuthal integrals. We now turn our attention to these.

The integral in \eqref{h_powerspec} depends on the azimuthal angles through the function $Q_{\lambda}$ as we just mentioned and also through the scalar 4-point function $\braket{\zeta^4}'$. The tensor power spectrum therefore has the form
\begin{equation}\label{Plambda phi int}
    P_{\lambda}=\int_0^{2\pi} \D \phi_1\int_0^{2\pi}\D\phi_2 \,\widetilde{Q}_{\lambda}(\phi_1)\widetilde{Q}^{*}_{\lambda}(\phi_2)\mathcal F(\phi_1-\phi_2),
\end{equation}
where
\begin{equation}
    \widetilde{Q}_{\lambda}(\phi)=\begin{cases}
    \cos2\phi\quad \lambda=+\\
    \sin2\phi\quad \lambda=\times \\
        \frac{1}{\sqrt{2}}e^{2i\phi}\quad \lambda=R\\
        \frac{1}{\sqrt{2}}e^{-2i\phi}\quad \lambda=L \\
    \end{cases}.
\end{equation}
The fact that the function $\mathcal{F}$ must depend only on the difference $\phi_1-\phi_2$ is a consequence of the rotational invariance of $\braket{\zeta^4}'$. 
In addition, because $\braket{\zeta^4}'$ must also of course be periodic in $\phi_{1,2}$, it follows that $\mathcal{F}$ is a periodic function with period $2\pi$.

Let us consider first the integral over the disconnected part of the scalar 4-point function. In this case the only dependence of $\braket{\zeta^4}_{\rm d}'$ on the azimuthal angles is through $\delta^{3}(\bm q_1-\bm q_2)$ and $\delta^3(\bm q_1+\bm q_2-\bm k)$, which in fact contribute equally due to the symmetries of the integral. Since $\delta^{3}(\bm q_1-\bm q_2)\propto\delta(\phi_1-\phi_2)$ we have that $\mathcal{F}=\widetilde{\mathcal{F}}\delta(\phi_1-\phi_2)$ with $\widetilde{\mathcal{F}}$ independent of $\phi_{1,2}$. Therefore
\begin{equation}
    P_{\lambda}=\widetilde{\mathcal{F}}\int_0^{2\pi} \D \phi_1\,|\widetilde{Q}_{\lambda}(\phi_1)|^2 = \pi\widetilde{\mathcal{F}} \qquad \forall\; \lambda\in\{+,\times,R,L\} \,.
\end{equation}
The total power spectrum, computed either in the linear basis or in the chiral basis, is thus equal to $P_h=2\pi\widetilde{\mathcal{F}}$. Since each polarization channel contributes equally, we conclude as expected that the disconnected SIGW power spectrum cannot exhibit parity violation.

Next we consider the connected 4-point function. Performing the change of variables
\begin{equation}
    \chi=\phi_1+\phi_2,\quad \psi=\phi_1-\phi_2,
\end{equation}
allows us to recast the azimuthal integrals as
\begin{equation}
    P_{+,\times}=\pi\int_0^{2\pi}\D \psi \,\cos2\psi\,\mathcal F(\psi),
\end{equation}
\begin{equation}
    P_R=\pi\int_0^{2\pi}\D \psi \,e^{2i\psi}\,\mathcal F(\psi), \qquad P_L=\pi\int_0^{2\pi}\D \psi \,e^{-2i\psi}\,\mathcal F(\psi),
\end{equation}
where we made use of the symmetry properties of the function $\mathcal{F}$.

We reach the conclusion that linearly polarized SIGWs always contribute equally to the total tensor power spectrum. This is simply a consequence of rotational invariance since $+$ and $\times$ waves are related by a rotation of $\pi/4$. The total power induced by the trispectrum is then given by
\begin{equation} \label{eq:totalP azimuthalint}
    P_h=2\pi\int_0^{2\pi}\D \psi \,\cos2\psi\,\mathcal F(\psi).
\end{equation}
Since $\mathcal{F}$ is periodic it may be expanded in Fourier series, $\mathcal{F}=\mathcal{F}_0+\sum_{n\geq1}c_n\cos n\psi+s_n\sin n\psi$, so that the integral in \eqref{eq:totalP azimuthalint} is non-zero only if the Fourier coefficient $c_2$ is non-zero. In particular, $\mathcal{F}(\psi)$ cannot be an odd function of $\psi$ for $P_h$ to be non-zero.

To be more specific, now recall that $\mathcal{F}$ is built out of the three momenta $\bm k$, $\bm q_1$ and $\bm q_2$, and rotational invariance dictates that it must be a function of the scalars
\begin{equation} \label{eq:rotational invariants}
    k,\quad q_1,\quad q_2,\quad {\bm k}\cdot{\bm q_1},\quad {\bm k}\cdot{\bm q_2},\quad {\bm q_1}\cdot{\bm q_2},\quad {\bm k}\cdot{\bm q_1\times \bm q_2}.
\end{equation}
The first five are independent of $\psi=\phi_1-\phi_2$, while
\begin{equation}
    {\bm q_1}\cdot{\bm q_2}\propto \cos\psi ,\qquad {\bm k}\cdot{\bm q_1\times \bm q_2}\propto \sin\psi.
\end{equation}
Thus we may write, in the case of trispectrum-induced GWs, $\mathcal{F}(\psi)=\mathcal{G}(\cos\psi,\sin\psi)$,\footnote{Obviously the definition of $\mathcal{G}$ is ambiguous since we could write $\sin^2\psi=1-\cos^2\psi$. This is a consequence of the fact that the invariant $({\bm k}\cdot{\bm q_1\times \bm q_2})^2$ is not independent of the other terms in \eqref{eq:rotational invariants}. We fix this ambiguity by assuming that higher powers of $({\bm k}\cdot{\bm q_1\times \bm q_2})$ have been reduced in this way, i.e.\ so that one is left either without such term or with a single power. \label{footnoteref}} and so the above integral may be recast as
\begin{equation}
    P_h=2\pi\int_0^{\pi}\D \psi \,\cos2\psi\left[\mathcal{G}(\cos\psi,\sin\psi)+\mathcal{G}(\cos\psi,-\sin\psi)\right].
\end{equation}
Thus we deduce the stronger statement that the trispectrum cannot be an odd function of the scalar product $({\bm k}\cdot{\bm q_1\times \bm q_2})$ as a necessary condition for $P_h$ to be non-vanishing. Since a strictly parity-odd trispectrum must be an odd function of $({\bm k}\cdot{\bm q_1\times \bm q_2})$, we conclude that any parity-odd component will vanish in the calculation of the total power spectrum. Moreover, because of the identity
\begin{equation}
    \int_0^{\pi}\D \psi \,\cos2\psi\,\mathcal{G}(\cos\psi,-\sin\psi)=\int_0^{\pi}\D \psi \,\cos2\psi\,\mathcal{G}(-\cos\psi,-\sin\psi),
\end{equation}
it also follows that, if the trispectrum is even in $({\bm k}\cdot{\bm q_1\times \bm q_2})$, then it cannot be odd in the scalar $({\bm q_1}\cdot{\bm q_2})$. In other words, a parity-even $\mathcal{T}_{\zeta}$ cannot be an odd function of $({\bm q_1}\cdot{\bm q_2})$ as a necessary condition to have a non-zero total GW power spectrum.\footnote{This condition is necessary but clearly not sufficient, as illustrated by the trivial case when $\mathcal{F}$ is independent of $\psi$.} If we assume that $\mathcal{T}_{\zeta}$ is a polynomial in these two scalar quantities, and given that even powers of $({\bm k}\cdot{\bm q_1\times \bm q_2})$ are redundant, we then infer that only monomials containing even powers of $({\bm q_1}\cdot{\bm q_2})$ can in principle yield a net non-zero contribution.

Unlike for linear polarizations, GWs in the chiral basis do not give equal contributions to the total power. From the above expressions for the azimuthal integrals we have
\begin{equation}
\begin{aligned}
  P_R-P_L&=2\pi i  \int_0^{2\pi}\D \psi \,\sin2\psi \, \mathcal F(\psi) \\
  &=2\pi i\int_0^{\pi}\D \psi \,\sin2\psi\left[\mathcal{G}(\cos\psi,\sin\psi)+\mathcal{G}(-\cos\psi,-\sin\psi)\right].
  \end{aligned}
\end{equation}
The first equality tells us that $\mathcal{F}$ cannot be an even function of $\psi$ for $P_R-P_L$ to be non-zero. It therefore must be an odd function of the scalar $({\bm k}\cdot{\bm q_1\times \bm q_2})$, as already mentioned (see footnote \ref{footnoteref}), and so $\mathcal{G}$ must be odd in its second argument (assuming a strictly parity-odd trispectrum). But the second equality shows that, if this is the case, then $\mathcal{G}$ cannot be even in its first argument.

It follows that a parity violating SIGW signal, as measured by the difference $P_R-P_L$, requires the trispectrum to be an odd function of $({\bm k}\cdot{\bm q_1\times \bm q_2})$ and \textit{not} an even function of $({\bm q_1}\cdot{\bm q_2})$. Once again, this is a necessary but not sufficient condition. It seems difficult to sharpen this criterion without further knowledge about the structure of the trispectrum, so to proceed let us again make the additional assumption that $\mathcal{T}_{\zeta}$ depends polynomially on $({\bm q_1}\cdot{\bm q_2})$ and $({\bm k}\cdot{\bm q_1\times \bm q_2})$. In this situation it is easy to see that the trispectrum must contain at least one odd power of\footnote{Non-polynomial functions of these scalars do not need to satisfy this criterion. To give an artificial example, the function $\mathcal{G}(\cos\psi,\sin\psi)=\sin(\cos\psi)\sin(\sin\psi)$ is odd in both of its arguments and $\int_0^{2\pi}\D\psi\,\sin2\psi\,\mathcal{G}\neq0$, yet it is not odd in the product $\sin\psi\cos\psi$.}
\begin{equation} \label{eq:product PO}
    (\bm k\cdot\bm q_1\times \bm q_2)(\bm q_1\cdot\bm q_2)\propto \sin2\psi,
\end{equation}
where the coefficients of such terms may depend on any of the other scalars in Eq.\ \eqref{eq:rotational invariants} which do not affect the azimuthal dependence. We will see in the next section that the parity-odd trispectra from Ghost Inflation are precisely linear in the combination in Eq.\ \eqref{eq:product PO} and hence satisfy the criterion to have $P_R\neq P_L$.

\subsection{Perturbativity bounds on trispectrum-induced GWs} \label{sec:pert bound}

In this subsection we recall the argument of \cite{Garcia-Saenz:2022tzu} that shows that the ratio of connected to disconnected contributions to the GW power spectrum is bounded by the requirement of perturbativity. This is perfectly intuitive since the connected component is precisely a measure of the interactions of the underlying theory. The essence of the argument is to make this statement more precise, by noting that
\begin{equation} \label{eq:perturbativity1}
\frac{\mathcal P_{h,\rm c}}{\mathcal P_{h,\rm d}}\sim \frac{\mathcal P_\zeta^{\rm(1-loop)}}{\mathcal P_\zeta^{\rm(tree)}} \,,
\end{equation}
i.e.\ the ratio of the 1-loop correction to the scalar power spectrum relative to its tree level result. We note that the ``$\sim$'' here means that we are performing a simple dimensional analysis and need not be an accurate order-of-magnitude estimate. We will return to this point at the end of the section.

To check the above statement, we first observe that
\begin{equation} \label{eq:perturbativity2}
\frac{\mathcal P_{h,\rm c}}{\mathcal P_{h,\rm d}}\sim \frac{\mathcal T_\zeta}{\mathcal P_\zeta^2} \,,
\end{equation}
as follows from Eqs.\ \eqref{Ph_disconnected} and \eqref{Ph_connected}, again ignoring all numerical factors. Consider first a cubic interaction vertex with Hamiltonian $H_{\rm int}$. Both the scalar-exchange diagram for the trispectrum and the 1-loop diagram for the scalar power spectrum contain two such vertices. In the first case we schematically have
\begin{equation}
    \mathcal{T}_{\zeta}\sim \left\langle \left[\int H_{\rm int}(\eta_1)\right]\zeta^4\left[\int H_{\rm int}(\eta_2)\right]\right\rangle\sim \mathcal P_\zeta^2\left[\int H_{\rm int}\right]^2,
\end{equation}
while for the 1-loop diagram
\begin{equation}
    \mathcal{P}_{\zeta}^{\rm(1-loop)}\sim \left\langle
    \left[\int 
    H_{\rm int}(\eta_1)
    \right]\zeta^2\left[\int
    H_{\rm int}(\eta_2)
    \right]
    \right\rangle\sim \mathcal P_\zeta\left[
    \int H_{\rm int}
    \right]^2.
\end{equation}
Therefore
\begin{equation}
\frac{\mathcal T_\zeta}{\mathcal P_\zeta^2}\sim \frac{\mathcal P_\zeta^{\rm(1-loop)}}{\mathcal P_\zeta^{\rm(tree)}}\sim \left[\int
H_{\rm int}
\right]^2,
\end{equation}
implying in particular the relation in \eqref{eq:perturbativity1}. It is easy to see that the same conclusion follows for a quartic interaction vertex, in which case both the corresponding contact diagram for the trispectrum and the loop diagram for the power spectrum contain a single insertion of the interaction Hamiltonian.

The condition of perturbative control dictates that a 1-loop correction, in this case of the scalar power spectrum, must be parametrically smaller than its tree-level value. Eq.\ \eqref{eq:perturbativity1} would therefore seem to bound any trispectrum-induced GW signal to be subleading. However, let us emphasize again that this relation is a result of naive dimensional analysis and there are a few reasons why this estimate may be incorrect: (i) the estimate in Eq.\ \eqref{eq:perturbativity2} involves a complicated integral and it is far from obvious whether the result must be of order unity; (ii) the above argument ignored the ever-present factor of $(4\pi)^{-2}$ that accompanies any loop integral; (iii) on the opposite direction, loop corrections may be also enhanced if a large number of light fields run in the loops.

In the situation of point (iii) the perturbativity bound becomes stronger, as indeed it was explicitly proved to be the case for local-type trispectra in Ref.\ \cite{Garcia-Saenz:2022tzu}. On the other hand, since reasons (i) and (ii) could a priori relax the bound, it is critical to search for models in which a large trispectrum-induced signal may be in principle obtained, which is precisely one of our motivations to study the scenario of Ghost Inflation.

%%%%%%%%%%%%%%%%%%%%%%%%%%%%%%%%%%%%%
%%%%%%%%%%%%%%%%%%%%%%%%%%%%%%%%%%%%%

\section{Induced GWs from Ghost Inflation}\label{sec:GI}

\subsection{Ghost Inflation}

Ghost Inflation is a scenario of the pre-Big Bang universe in which a de Sitter phase of expansion arises from the condensation of ghost-type scalar field $\varphi$. The vacuum expectation value corresponds to a superfluid phase, i.e.\ a time-dependent background given by
\begin{equation}
\braket{\varphi}=M^2t ,
\end{equation}
where $M$ is an energy scale. The Lagrangian for $\varphi$ is Lorentz invariant and also invariant under the internal shift $\varphi\mapsto \varphi+c\,$. The background $\braket{\varphi}$ spontaneously breaks time translations and the shift symmetry, but preserves a linear combination of these two.

The Nambu-Goldstone boson $\pi(t,\bm x)$, or ``ghostone'' in this context, associated to the broken part of the symmetry group corresponds to the fluctuation of the scalar field about its vacuum, i.e.\ $\varphi=M^2t+\pi$, and inherits the shift symmetry of the microscopic theory, $\pi\mapsto \pi+c\,$.

Ghost Inflation has two peculiar features that distinguish it from more generic single-field models of inflation. The first is that the shift symmetry of the ghostone is \textit{exact}, at least in its minimal formulation. The second is that the speed of sound of the ghostone quanta vanishes, i.e.\ the ghostone has no standard two-derivative gradient energy. The leading gradients in the quadratic Lagrangian for $\pi$ are then given by a four-derivative operator,
\begin{equation}\label{action_ghost}
    S_{\pi^2}=\int \D\eta\D^3x~\frac{a^4}{2}\left[
    \frac{\pi'^2}{a^2}-\frac{\alpha}{M^2}\frac{(\partial^2\pi)^2}{a^4}
\right].
\end{equation}
Here $\alpha$ is a constant parameter that we may take to be of order unity without loss of generality. Let us also remark that this is not the full quadratic action for the perturbations as two simplifications have been made. The first is that couplings to the metric fluctuation have been neglected at this order, which may be shown to be a valid approximation provided $H\lesssim M\ll M_{\rm Pl}$. The second simplification is that certain ghostone terms have been dropped, which is valid under the assumption $H\ll M$; see \cite{Izumi:2010wm} for details. Under these assumptions, we see that the ghostone exhibits a non-linear dispersion relation given by
\begin{equation} \label{eq:disp relation}
    \omega^2=\frac{\alpha}{M^2}\frac{k^4}{a^4}.
\end{equation}

Solving the linear EOM in Fourier space, and imposing that in the asymptotic past the field is in its Minkowski vacuum, one finds the mode function
\begin{equation}\label{pi_solu}
    \pi_k(\eta)=H\sqrt{\frac{\pi}{8}}(-\eta)^{3/2}H_{3/4}^{(1)}(q\eta^2),\qquad\qquad  q\equiv \frac{\sqrt\alpha Hk^2}{2M}.
\end{equation}
Here $H_{\nu}^{(1)}(z)$ is the Hankel function of the first kind. Recall that the comoving curvature perturbation $\zeta$ is related to the ghostone $\pi$ at linear order via \cite{Maldacena:2002vr} 
\begin{equation}
    \zeta=-\frac{H}{\braket{\dot{\varphi}}}\pi=-\frac{H}{M^2}\pi.
\end{equation}
The dimensionless power spectrum of $\zeta$ is defined by
\begin{equation}
    \mathcal P_{\zeta}(k)=\frac{k^3}{2\pi^2}P_\zeta(k),\qquad \qquad \braket{\zeta_{\bm k}\zeta_{\bm k'}}=(2\pi)^3\delta(\bm k+\bm k')P_\zeta(k),
\end{equation}
and so we obtain the following expression for the primordial scalar power spectrum in Ghost Inflation:
\begin{equation}
    \mathcal P_\zeta=\frac{\alpha^{-3/4}}{\pi \Gamma^2(1/4)}\left(
    \frac{H}{M}\right)^{5/2}.
\end{equation}
Once again, here we assume exact scale invariance so that $H$, and hence $\mathcal{P}_{\zeta}$, is supposed constant and independent of $k$.

Let us next consider self-interactions of the ghostone $\pi$. One first observes that the non-relativistic dispersion relation in Eq.\ \eqref{eq:disp relation} implies an unusual power counting scheme in order to determine the most relevant couplings. Taking into account the shift symmetry, and assuming in addition a $\mathbb Z_2$ symmetry $\varphi\mapsto -\varphi$ in the ultraviolet theory, it may be shown that the leading operators in the effective field theory for the ghostone are given by\cite{Arkani-Hamed:2003juy,Izumi:2010wm}
\begin{align}
    S_{\pi^3}&=-\frac{\beta}{2M^2}\int\D\eta\D^3x\, a \pi'\partial_i\pi\partial_i\pi,\label{Action_cubic}\\
    S_{\pi^4}&=-\frac{\gamma}{8M^4}\int \D\eta\D^3x (\partial_i\pi\partial_i\pi)^2\label{Action_quartic_1}.
\end{align}
Here $\beta$ and $\gamma$ are dimensionless coupling constants, expected to be of order unity on the basis of naturalness. Here however we will not make any assumptions regarding their size.

The previous operators are even under a parity inversion, $\pi(\eta,\bm x)\mapsto\pi(\eta,-\bm x)$. In this paper we are interested instead in exploring the effects of a parity-breaking scalar Lagrangian, which leads us to consider the following quartic interactions \cite{Cabass:2022rhr}:
\begin{align} \label{eq:PO lagrangians1}
    S_{\pi^4}^{\rm PO,1}&=\frac{\gamma_1}{8M^{10}}\int \D\eta\D^3x \,a^{-6}\pi'\epsilon_{ijk}\partial_{il}\pi\partial_{lj}\partial^2\pi\partial_k\partial^2\pi,\\
    S_{\pi^4}^{\rm PO,2}&=\frac{\gamma_2}{8M^9}\int \D\eta\D^3x\, a^{-5}\epsilon_{ijk}\partial_{mn}\pi\partial_{ni}\pi\partial_{mlj}\pi\partial_{lk}\pi,\label{eq:PO lagrangians2}
\end{align}
where $\partial_{ij\cdots}\equiv \partial_i\partial_j\partial_{\cdots}$ and $\gamma_{1,2}$ are dimensionless coefficients. These terms contain an odd number of spatial derivatives and hence manifestly break parity. The anti-symmetric contraction of indices will lead, as we will see explicitly, to cross products of the momenta in the 4-point function, again making the breaking of parity manifest, $\bm k\mapsto -\bm k$ in Fourier space. An additional consequence of the violation of parity is that the 4-point function for these couplings will turn out to be purely imaginary as is well known. Finally let us mention that the above list is of course not exhaustive but may be shown to give the most relevant parity-odd terms (in the sense of the derivative expansion) \cite{Cabass:2022oap}.

\subsection{Power spectrum-induced GWs}

Since Ghost Inflation predicts a scale invariant scalar power spectrum, the calculation of the master integral for the disconnected contribution to the GW spectrum reduces to the same integral that one would find in any scale-invariant scenario. As first computed in \cite{Kohri:2018awv} we verify that the result is
\begin{equation} \label{eq:Ph disc}
\overline{\mathcal P_{h,\rm d}}=\mathcal{I}_{\rm d} \left(\frac{\mathcal H}{k}\right)^2\mathcal P_\zeta^2 , \qquad \mathcal{I}_{\rm d}\approx 9.87 ,
\end{equation}
where $\mathcal{I}_{\rm d}$ was obtained after a numerical calculation of the master integral in Eq.\ \eqref{Ph_disconnected}. See Appendix \ref{sec:app numerics} for details regarding our numerical computations.

\subsection{Parity-even trispectrum-induced GWs} \label{sec:PE GW}

In this subsection we calculate the GW power spectrum induced by the trispectra that derive from the parity-even (PE) operators in Eqs.\ \eqref{Action_cubic} and \eqref{Action_quartic_1}. We compare the results with the above Gaussian contribution and estimate bounds on the size of the ratio from the requirement of perturbativity.

\subsubsection*{Scalar-exchange diagram}

The cubic coupling in Eq.\ \eqref{Action_cubic} contributes to the scalar trispectrum through an exchange diagram with two insertions of the interaction Hamiltonian. The result is \cite{Izumi:2010wm,Huang:2010ab}
\begin{equation}
\begin{aligned}
\mathcal{T}_{\zeta}(\bm k_1,\bm k_2,\bm k_3,\bm k_4)&=\frac{1}{2^{9}\pi\Gamma^4(1/4)}\left(\frac{H}{M}\right)^{11}\frac{\beta^2}{\alpha^{1/2}}( k_1 k_2 k_3 k_4)^{3/4} \\
&\quad\times\left[-2J_{1A}(\bm k_1,\bm k_2,\bm k_3,\bm k_4)+J_{1B}(\bm k_1,\bm k_2,\bm k_3,\bm k_4)+(23~\mathrm{perm.})\right],
\end{aligned}
\end{equation}
where
\begin{equation}
\begin{aligned}
\,& J_{1A}(\bm k_1,\bm k_2,\bm k_3,\bm k_4)={\rm Re}\bigg[\int_{-\infty}^0\D\eta(-\eta)^{9/2}\int_{-\infty}^{\eta}\D\eta'(-\eta')^{9/2} \\
&\times\bigg((\bm k_1\cdot\bm k_2)(k_1^2+k_2^2+2\bm k_1\cdot\bm k_2)H_{3/4}^{(1)}(q_1\eta^{\prime2})H_{3/4}^{(1)}(q_2\eta^{\prime2})H_{-1/4}^{(1)}(q_{12}\eta^{\prime2}) \\
&\quad -2(k_1^2+\bm k_1\cdot\bm k_2)k_1^2H_{3/4}^{(1)}(q_2\eta^{\prime2})H_{-1/4}^{(1)}(q_1\eta^{\prime2})H_{3/4}^{(1)}(q_{12}\eta^{\prime2})\bigg) \\
&\times\bigg((\bm k_3\cdot\bm k_4)(k_3^2+k_4^2+2\bm k_3\cdot\bm k_4)H_{3/4}^{(1)}(q_3\eta^{2})H_{3/4}^{(1)}(q_4\eta^{2})H_{-1/4}^{(1)}(q_{34}\eta^{2}) \\
&\quad -2(k_3^2+\bm k_3\cdot\bm k_4)k_4^2H_{3/4}^{(1)}(q_3\eta^{2})H_{-1/4}^{(1)}(q_4\eta^{2})H_{3/4}^{(1)}(q_{34}\eta^{2})\bigg)\bigg], \\
\end{aligned}
\end{equation}
where $q_i\equiv \frac{\sqrt{\alpha}H k_i^2}{2M}$ and $q_{ij}\equiv \frac{\sqrt{\alpha}H({\bm k}_i+{\bm k}_j)^2}{2M}$. The integral $J_{1B}$ has exactly the same integrand with the difference that the upper limit of the second integral is $\eta'=0$, and notice that in this case the integral is real; see Appendix \ref{sec:app recast integrals} for details.

From the master integral for the connected SIGW spectrum we derive
\begin{equation} \label{eq:Ph PE1}
\overline{\mathcal P_{h,\rm PE1}}=\mathcal{I}_{\rm PE1}\left(\frac{\mathcal H}{k}\right)^2\frac{\beta^2}{\alpha^{8/5}}\mathcal P_\zeta^{11/5} , \qquad \mathcal{I}_{\rm PE1}\approx {-2.4\times 10^{-3}}.
\end{equation}
This result is seemingly suppressed relative to the power spectrum-induced result, Eq.\ \eqref{eq:Ph disc}, as it has a smaller numerical coefficient and a larger power of $\mathcal{P}_{\zeta}$. Recall that we are entertaining the possibility of having  a strong enhancement of $\mathcal{P}_{\zeta}$ relative to its value on CMB scales, yet we still expect $\mathcal{P}_{\zeta}\ll 1$ for the background not to be spoiled by backreaction. Nevertheless, we see that $\overline{\mathcal P_{h,\rm PE1}}$ depends also on the coupling constants $\alpha$ and $\beta$, and one could a priori consider the situation where the particular ratio in Eq.\ \eqref{eq:Ph PE1} is large, even if this would entail a fine-tuning of the theory. However, based on the discussion of Sec.\ \ref{sec:pert bound}, we anticipate that a large non-Gaussian contribution must be in tension with the condition of perturbative control.

To verify this we first divide this result by the disconnected contribution,
\begin{equation}
\frac{\overline{\mathcal P_{h,\rm PE1}}}{\overline{\mathcal P_{h,\rm d}}}=\frac{\mathcal{I}_{\rm PE1}}{\mathcal{I}_{\rm d}} \frac{\beta^2}{\alpha^{8/5}}\mathcal P_\zeta^{1/5}=\mathcal{C}_{\rm PE1}\frac{\beta^2}{\alpha^{7/4}}\left(\frac{H}{M}\right)^{1/2},\qquad \mathcal{C}_{\rm PE1}\approx {-1.2\times 10^{-4}}.
\end{equation}
and we wish to compare this ratio to the relative 1-loop correction to the scalar power spectrum, as discussed in Sec.\ \ref{sec:pert bound}.

In fact, in the case of the scalar-exchange term, we can draw from the analysis of \cite{Baumann:2011su} a more precise estimate of the perturbativity bound, namely the strong coupling scale that follows from the requirement of perturbative unitarity,\footnote{To make contact with the result of \cite{Baumann:2011su}, we have the following dictionary relating their parameters to the ones of the Ghost Inflation Lagrangian: $\rho^{-2}=\alpha/M^2$, $M_2^2=M^2/(2\beta)$. Note that the field $\pi$ in \cite{Baumann:2011su} is not canonically normalized as it is in \eqref{action_ghost}.}
\begin{equation}
    \Lambda_*=4\pi^2M\alpha^{7/2}\beta^{-4}.
\end{equation}
For the Ghost Inflation model to be a valid effective description we demand that $H\ll \Lambda_*$, that is
\begin{equation}
\left(\frac{H}{\Lambda_*}\right)^{1/2}=\frac{1}{2\pi}\frac{\beta^2}{\alpha^{7/4}}\left(\frac{H}{M}\right)^{1/2}\ll 1,
\end{equation}
so that
\begin{equation}
\frac{\overline{\mathcal P_{h,\rm PE1}}}{\overline{\mathcal P_{h,\rm d}}}\ll 2\pi\mathcal{C}_{\rm PE1}= \mathcal{O}(10^{-3}).
\end{equation}
This result confirms the no-go theorem of \cite{Garcia-Saenz:2022tzu}, in fact with a bound that is significantly stronger than the one obtained in standard slow-roll inflation or in models with reduced speed of sound.

\subsubsection*{Contact diagram}

Another contribution to the scalar trispectrum is given by a contact diagram with a single insertion of the quartic interaction Hamiltonian due to the coupling in Eq.\ \eqref{Action_quartic_1}.\footnote{Here and below we assume for simplicity that the cubic vertex of Eq.\ \eqref{Action_cubic} is turned off when calculating the trispectrum. Otherwise this cubic Lagrangian also contributes to the quartic interaction Hamiltonian. The resulting expression is however the same, only with the change $\gamma\to \tilde{\gamma}=\gamma+\beta^2$ \cite{Huang:2010ab}.} The resulting trispectrum function is given by \cite{Huang:2010ab,Izumi:2010wm}
\begin{equation}
\begin{aligned}
\mathcal{T}_{\zeta}(\bm k_1,\bm k_2,\bm k_3,\bm k_4)&=\frac{1}{2^{12}\pi^2\Gamma^4(1/4)}\left(\frac{H}{M}\right)^{9}\frac{\gamma}{\alpha^{3/2}}\,(k_1k_2k_3k_4)^{3/4}J_2(k_1,k_2,k_3,k_4) \\
&\quad \times\left[(\bm k_1\cdot\bm k_2)(\bm k_3\cdot\bm k_4)+(23~\mathrm{perm.})\right],
\end{aligned}
\end{equation}
where
\begin{equation}
J_2(k_1,k_2,k_3,k_4)={\rm Re}\left[i\int_{-\infty}^0\D\eta\,\eta^6H_{3/4}^{(1)}(q_1\eta^2)H_{3/4}^{(1)}(q_2\eta^2)H_{3/4}^{(1)}(q_3\eta^2)H_{3/4}^{(1)}(q_4\eta^2)\right],
\end{equation}
and $q_i\equiv \frac{\sqrt{\alpha}H k_i^2}{2M}$.

Performing the numerical calculation of the master integral we obtain
\begin{equation}
    \overline{\mathcal P_{h,\rm PE2}}=\mathcal{I}_{\rm PE2} \left(\frac{\mathcal H}{k}\right)^2\frac{\gamma}{\alpha^{8/5}}\mathcal P_\zeta^{11/5},\qquad \mathcal{I}_{\rm PE2}\approx {-2.0\times 10^{-2}}.
\end{equation}
The ratio of this outcome with the power spectrum-induced result is then
\begin{equation} \label{eq:PE2 ratio}
    \frac{|\overline{\mathcal P_{h,\rm PE2}}|}{\overline{\mathcal P_{h,\rm d}}}= \frac{|\mathcal{I}_{\rm PE2}|}{\mathcal{I}_{\rm d}}\frac{\gamma}{\alpha^{8/5}}\mathcal P_\zeta^{1/5}=\mathcal{C}_{\rm PE2}\frac{\gamma}{\alpha^{7/4}}\left(\frac{H}{M}\right)^{1/2}, \qquad \mathcal{C}_{\rm PE2}\approx {9.6\times 10^{-4}}.
\end{equation}

We again wish to compare this with the bound on the 1-loop power spectrum dictated by perturbativity. In this case, i.e.\ in the situation when the cubic operator in Eq.\ \eqref{Action_cubic} has been artificially turned off (see the comment in the footnote), the precise unitarity bound corresponding to the quartic coupling has not been computed. Nevertheless we may still get an order-of-magnitude estimate simply from the knowledge of the form of the interaction Hamiltonian, as explained in Sec.\ \ref{sec:pert bound}. We assume the 1-loop result is suppressed by a factor of $(4\pi)^{-2}$ relative to what dimensional analysis would predict, i.e.
\begin{equation}
    \frac{\mathcal P_\zeta^{\rm(1-loop)}}{\mathcal P_\zeta^{\rm(tree)}}\sim \frac{1}{(4\pi)^2}\frac{\gamma}{\alpha^{7/4}}\left(
\frac{H}{M}
\right)^{1/2}\ll 1.
\end{equation}
It follows that
\begin{equation}
\frac{|\overline{\mathcal P_{h,\rm PE2}}|}{\overline{\mathcal P_{h,\rm d}}} \ll (4\pi)^2\mathcal{C}_{\rm PE2}=\mathcal{O}(10^{-1}).
\end{equation}
This result is similar to the bound obtained in standard models of inflation with reduced speed of sound \cite{Garcia-Saenz:2022tzu} and implies a strong suppression of the relative importance of the trispectrum-induced GW signal.

\subsection{Parity-odd trispectrum-induced GWs} \label{sec:PO GW}

\subsubsection*{First diagram}

Next we consider the interaction Hamiltonian produced by the first parity-odd (PO) term in Eq.\ \eqref{eq:PO lagrangians1}. The trispectrum reads \cite{Cabass:2022rhr}
\begin{equation}
\begin{aligned}
     \mathcal{T}_{\zeta}(\vec k_1,\vec k_2,\vec k_3,\vec k_4)=&\;\frac{i(2\pi)^3}{\Gamma(3/4)^2}\left(\frac{H}{M}\right)^{3/2}\frac{\gamma_1}{\alpha^{9/4}}\mathcal{P}_\zeta^3 \\
     &\times\left[(\vec k_2\cdot \vec k_3\times\vec k_4)(\vec k_2\cdot \vec k_3)\frac{(k_1k_2k_3k_4)^{11/4}}{k_2^2}I_1(k_1,k_2,k_3,k_4)+(23~\mathrm{perm.})\right],
\end{aligned}
\end{equation}
where
\begin{equation}
I_1(k_1,k_2,k_3,k_4)=\int _0^\infty d\lambda \lambda^{13}H_{-1/4}^{(1)}(2ik_1^2\lambda^2)H_{3/4}^{(1)}(2ik_2^2\lambda^2)H_{3/4}^{(1)}(2ik_3^2\lambda^2)H_{3/4}^{(1)}(2ik_4^2\lambda^2).
\end{equation}

The trispectrum involves the product $(\bm k_2\cdot \bm k_3\times \bm k_4)(\bm k_2\cdot \bm k_3)$, which upon replacing $\bm k_1=\bm q_1$, $\bm k_2=\bm k-\bm q_1$, $\bm k_3=-\bm q_2$, $\bm k_4=-\bm k+\bm q_2$ in the master integral yields a dependence on the azimuthal angle,
\begin{equation}
\bm k\cdot(\bm q_1\times \bm q_2)(\bm q_1\cdot \bm q_2)=-kq_1^2q_2^2\sin^2\theta_1\sin^2\theta_2\sin2\psi-\frac14kq_1^2q_2^2\sin2\theta_1\sin2\theta_2\sin\psi.
\end{equation}
Recall from our discussion in Sec.\ \ref{polar_SIGWs} that odd functions of $\bm k\cdot(\bm q_1\times \bm q_2)(\bm q_1\cdot \bm q_2)$ may in principle induce a parity asymmetry between right- and left-handed polarizations, while it is guaranteed to yield identically zero in the total GW power.

The master integral for the difference of right- and left-handed SIGW spectra is
\begin{equation}
\begin{aligned}
        \overline{\mathcal P_{h,R}}-\overline{\mathcal P_{h,L}}= &\; i \left(\frac{\mathcal H}{k}\right)^2 \int_0^{\infty} \mathrm{d} v_1 \int_{\left|1-v_1\right|}^{1+v_1} \mathrm{~d} u_1 \int_0^{\infty} \mathrm{d} v_2 \int_{\left|1-v_2\right|}^{1+v_2} \mathrm{~d} u_2 \int_0^{2 \pi} \mathrm{d} \psi \,\mathcal{K}_{\mathrm{c}}\left(u_1, v_1, u_2, v_2\right)\label{Ph_PR-PL} \\
& \times \frac{\sin (2 \psi)}{\pi} \mathcal{T}_\zeta\left(u_1, v_1, u_2, v_2, \psi\right).
\end{aligned}
\end{equation}
Carrying out the numerical calculation we find 
\begin{equation}
\overline{\mathcal P_{h,R}}-\overline{\mathcal P_{h,L}}=\mathcal{I}_{\rm PO1}\frac{\gamma_1}{\alpha^{9/5}}\left(\frac{\mathcal H}{k}\right)^2\mathcal P_\zeta^{18/5} ,\qquad \mathcal{I}_{\rm PO1}\approx {-3.3\times 10^{-1}} .
\end{equation}
See again Appendices \ref{sec:app recast integrals}, \ref{sec:app numerics} for details on the numerics, in particular concerning how to recast the integrals in a form amenable to numerical integration.

The degree of parity violation may be quantified by the so-called chirality parameter \cite{Sato-Polito:2019hws,Gluscevic:2010vv,Zhang:2022xmm}
\begin{equation}
\Pi\equiv \frac{\overline{\mathcal P_{h,R}}-\overline{\mathcal P_{h,L}}}{\overline{\mathcal P_{h,{\rm tot}}}}.
\end{equation}
Note that in this formula $\overline{\mathcal P_{h,R}}$ and $\overline{\mathcal P_{h,L}}$ should be understood as the total chiral power spectra. Here we assume for simplicity that the first parity-odd quartic coupling in Eq.\ \eqref{eq:PO lagrangians1} is the only interaction, although by consistency one still needs to include the disconnected contribution in the calculation of the GW spectrum. The latter vanishes in the difference $\overline{\mathcal P_{h,R}}-\overline{\mathcal P_{h,L}}$, since as discussed in Sec.\ \ref{polar_SIGWs} disconnected SIGWs preserve parity, whereas on the other hand the trispectrum-induced corrections vanish upon summing over polarizations. Thus we obtain
\begin{equation}
\Pi=-\mathcal{C}_{\rm PO1}\frac{\gamma_1}{\alpha^3}\left(\frac{H}{M}\right)^4,\qquad  \mathcal{C}_{\rm PO1}\approx {8.8\times 10^{-5}}.
\end{equation}

We estimate the perturbativity bound derived from this quartic coupling as
\begin{equation}
    \frac{P_\zeta^{\rm(1-loop)}}{\mathcal P_\zeta^{\rm(tree)}}\sim \frac{1}{(4\pi)^2}\frac{\gamma_1}{\alpha^3}\left(\frac{H}{M}\right)^4\ll 1,
\end{equation}
so that
\begin{equation}
|\Pi|\ll (4\pi)^2\mathcal{C}_{\rm PO1}=\mathcal{O}(10^{-2}) \,.
\end{equation}

\subsubsection*{Second diagram}

Proceeding with the second parity-odd Lagrangian in Eq.\ \eqref{eq:PO lagrangians2}, the dimensionless trispectrum in this case is given by \cite{Cabass:2022rhr}
\begin{equation}
\begin{aligned}
\mathcal{T}_{\zeta}(\vec k_1,\vec k_2,\vec k_3,\vec k_4)=&\; \frac{2i\pi^3}{\Gamma(3/4)^2}\left(
\frac{H}{M}
\right)^{1/2}\frac{\gamma_3}{\alpha^{9/4}}\mathcal P_\zeta^3\,(k_1k_2k_3k_4)^{3/4}I_2(k_1,k_2,k_3,k_4) \\
&\times\left[(\vec k_2\cdot \vec k_3\times \vec k_4)(\vec k_1\cdot \vec k_3)(\vec k_1\cdot \vec k_2)(\vec k_3\cdot \vec k_4)+(23~\mathrm{perm.})\right],
\end{aligned}
\end{equation}
where
\begin{equation}
    I_2(k_1,k_2,k_3,k_4)={\rm Im}\left[\int _0^\infty \D\lambda~ \lambda^{11}H_{3/4}^{(1)}(2ik_1^2\lambda^2)H_{3/4}^{(1)}(2ik_2^2\lambda^2)H_{3/4}^{(1)}(2ik_3^2\lambda^2)H_{3/4}^{(1)}(2ik_4^2\lambda^2)\right].
\end{equation}
Upon substitution in the master integral, the only non-zero independent scalar from the products $\bm k_i\cdot \bm k_j\times \bm  k_k$ for all channels is $\bm k\cdot \bm q_1\times\bm  q_2$, again implying a vanishing contribution to the total GW power.
As for the products $(\vec k_1\cdot \vec k_3)(\vec k_1\cdot \vec k_2)(\vec k_3\cdot \vec k_4)$ and permutations, it is easy to see that for each channel this combination contains precisely one factor of $\bm q_1\cdot \bm q_2$, indicating that a non-zero difference between the tensor power spectrum in right- and left-handed polarizations.

Explicitly we obtain
\begin{equation}
\overline{\mathcal P_{h,R}}-\overline{\mathcal P_{h,L}}= \mathcal{I}_{\rm PO2}\frac{\gamma_2}{\alpha^{21/10}} \mathcal P_\zeta ^{16/5}\left(\frac {\mathcal{H}} {k} \right)^2 ,\qquad \mathcal{I}_{\rm PO2}\approx {-8.2\times 10^{-1}} ,
\end{equation}
and a corresponding chirality parameter
\begin{equation} \label{eq:PO2 chirality}
\Pi=-\mathcal{C}_{\rm PO2}\frac{\gamma_2}{\alpha^3}\left(\frac{H}{M}\right)^3,\qquad  \mathcal{C}_{\rm PO2}\approx {9.5\times10^{-4}}.
\end{equation}

Performing a rough estimate of the 1-loop scalar power spectrum as done before we get the bound
\begin{equation}
|\Pi|\ll (4\pi)^2\mathcal{C}_{\rm PO2}=\mathcal{O}(10^{-1}) \,.
\end{equation}
This is an order of magnitude less restrictive than the bound derived for the first parity-odd interaction. Nevertheless, in either case one concludes that a parity-violating signal in SIGWs is strongly restricted by the assumption of perturbative control.

%=======================================

\section{Conclusions} \label{sec:conc}

In this work we presented a first study of SIGWs produced by the non-Gaussian component of the scalar 4-point correlation function in a scenario that goes beyond standard scenarios of inflation described by the effective field theory with reduced speed of sound as well as models that predict dominant local-type non-Gaussianities. The scenario of Ghost Inflation is further motivated by the possibility of having large scalar fluctuations on the scales of interest to next-generation GW detectors and, most interestingly from a theoretical perspective, the fact that the model may in principle include a parity-violating component in the scalar 4-point function.

Scalar trispectrum-induced GWs are necessarily small relative to the power spectrum-induced ones, as recently formalized in the no-go theorem of \cite{Garcia-Saenz:2022tzu}. Our findings further strengthen this expectation by extending previous analyses to encompass Ghost Inflation. Our results should straightforwardly generalize to other models with modified dispersion relation, although it would be worthwhile to confirm this explicitly. We find that the ratio of connected to disconnected contributions to $\mathcal{P}_h$ is bounded by a number of $\mathcal{O}(10^{-1})$ for the contact parity-even diagram of Sec.\ \ref{sec:PE GW} and for the second parity-odd diagram of Sec.\ \ref{sec:PO GW}, while the other terms are even more suppressed. It should be remarked that these bounds are rather optimistic since they made use of a crude estimate of the 1-loop correction to the scalar power spectrum whereas a more precise calculation of the unitarity bound, like the one done for the scalar-exchange diagram, might yield a stronger constraint.

Our work should be seen as a first analysis on the possibility of having a violation of parity symmetry in SIGWs from inflation. This is arguably a more conservative expectation in comparison to exotic descriptions where parity is broken at the level of the gravitational sector during radiation domination, although it would be interesting to compare the predictions of both scenarios in detail. This as well as many other applications would necessarily require one to go beyond the assumption of exact scale invariance. On the one hand, a scale dependence in the scalar correlation functions would break the degeneracy between the connected and disconnected parts of the \textit{total} SIGW power spectrum. Although the parity-odd component is, on the other hand, already non-degenerate when measured in the \textit{individual} chiral polarization channels, even assuming exact scale invariance, a characteristic scale dependence would potentially make its detectability far more likely. While our bounds derived from perturbativity cannot immediately be extrapolated to situations with strong scale dependence, they may nevertheless still be expected to be qualitatively correct as argued in \cite{Garcia-Saenz:2022tzu}.

It is intriguing that the chirality parameter for the second parity-odd interaction, Eq.\ \eqref{eq:PO2 chirality}, may in principle be as large as the relative contribution to GWs from the parity-even contact term, Eq.\ \eqref{eq:PE2 ratio}, while maintaining perturbative control. It should be emphasized however that this is rather optimistic. Consider for simplicity a scenario where only these two interactions are present. We may compare the importance of the resulting SIGWs by considering the ratio
\begin{equation}
\left|\frac{\overline{\mathcal P_{h,R}}-\overline{\mathcal P_{h,L}}}{\overline{\mathcal P_{h,{\rm PE2}}}}\right| = \mathcal{O}(10^{-1})\frac{\gamma_2}{\alpha^{5/4}\gamma}\left(\frac{H}{M}\right)^{5/2}.
\end{equation}
This is suppressed both by a small numerical prefactor and by a power of $H/M$, which should be recalled must be small for the effective description of Ghost Inflation to be valid. For this ratio to be sizable one therefore needs the combination $\frac{\gamma_2}{\alpha^{5/4}\gamma}$ of coupling constants to be large. While this is not forbidden by perturbativity, at least as inferred from our rough estimates of the 1-loop scalar power spectrum, it would still require unnatural values of Wilson coefficients and hence a fine tuning of the theory. It would be interesting to assess this point more rigorously, and more ambitiously to explore the possibility of having effective theories of inflation with naturally large parity-violating interactions.

%=======================================

\subsubsection*{Acknowledgments}

We are grateful to Lucas Pinol, S\'ebastien Renaux-Petel, Denis Werth, Fengge Zhang and Tao Zhu for helpful comments on the manuscript. This work received support from the NSFC Research Fund for International Scientists (Grant No.\ 12250410250), the China Postdoctoral Science Foundation under Grant No.\ 2022TQ0140 and the National Natural Science Foundation of China under Grant No.\ 12247161.

%==========================

\appendix

\section{Different conventions}

The use of different conventions in the SIGW literature can make it difficult to compare results among different works. In this appendix we summarize some important formulae for the two common conventions used in the definition of $h_{ij}$. This leads to different numerical prefactors in the expressions for $\mathcal{P}_h$ although the observables $\rho_{\rm GW}$ and $\Omega_{\rm GW}$ are of course independent of conventions.

\subsection{$\delta g_{ij}=h_{ij}$}

First we consider the second-order tensor perturbation defined in the present paper in \eqref{h_def}.
This is the definition adopted for instance in Refs.\ \cite{Domenech:2021ztg,Garcia-Saenz:2022tzu}.
The energy density is defined by
\begin{equation}
\rho_{\rm GW}=\frac{M_{\rm Pl}^2}{4 a^2}\overline{\braket{
    h_{ij}'h_{ij}'
}}
\end{equation}
The energy density spectrum in this convention is
\begin{equation}
\Omega_{\rm GW}(k)=\frac{1}{3M_{\rm Pl}^2H^2}\frac{\D \rho_{\rm GW}}{\D \log k}=\frac{1}{12}\left(\frac{k}{\mathcal H}\right)^2\sum_\lambda\mathcal P_{\lambda}(k).
\end{equation}
The resulting master integrals for the disconnected and connected contributions are given in \eqref{Ph_disconnected} and \eqref{Ph_connected} respectively.

\subsection{$\delta g_{ij}=\frac12 h_{ij}$}

This convention is adopted for instance in Refs.\ \cite{Ananda:2006af,Baumann:2007zm,Kohri:2018awv,Adshead:2021hnm}. The GW energy density and energy density spectrum are defined by
\begin{equation}
\rho_{\rm GW}=\frac{M_{\rm Pl}^2}{16 a^2}\overline{\braket{
    h_{ij}'h_{ij}'}},\qquad \Omega_{\rm GW}(k)=\frac{1}{48}\left(\frac{k}{\mathcal H}\right)^2\sum_{\lambda}\mathcal P_\lambda(k).
\end{equation}
The integral expressions for the disconnected and connected contributions to the SIGWs are given by
\begin{align}
\overline{\mathcal{P}_{h,\mathrm{d}}}= &\;4 \left(\frac{\mathcal H}{k}\right)^2 \int_0^{\infty} \mathrm{d} v \int_{|1-v|}^{1+v} \mathrm{~d} u \,\mathcal{K}_{\mathrm{d}}(u, v) \mathcal{P}_\zeta(k u) \mathcal{P}_\zeta(k v), \\
\overline{\mathcal{P}_{h, \mathrm{c}}}= &\;4\left(\frac{\mathcal H}{k}\right)^2 \int_0^{\infty} \mathrm{d} v_1 \int_{\left|1-v_1\right|}^{1+v_1} \mathrm{~d} u_1 \int_0^{\infty} \mathrm{d} v_2 \int_{\left|1-v_2\right|}^{1+v_2} \mathrm{~d} u_2 \int_0^{2 \pi} \mathrm{d} \psi\, \mathcal{K}_{\mathrm{c}}\left(u_1, v_1, u_2, v_2\right) \\
& \times \frac{\cos (2 \psi)}{\pi} \mathcal{T}_\zeta\left(u_1, v_1, u_2, v_2, \psi\right)\notag,
\end{align}
which multiply by $4$ those given in \eqref{Ph_disconnected} and \eqref{Ph_connected}.

\section{Recasting the integrals} \label{sec:app recast integrals}

\subsection{Parity-even scalar-exchange term}

The scalar-exchange contribution to the 4-point function is given in the in-in formalism as
\begin{equation}\label{trispectrum_SE}
    \begin{aligned}
        &\braket{\pi_{\bm k_1}(\eta_0)\pi_{\bm k_2}(\eta_0)\pi_{\bm k_3}(\eta)\pi_{\bm k_4}(\eta_0)}\\
        =&2\Re{-\int_{-\infty}^{\eta_0}\D\eta _2\int_{-\infty}^{\eta_2}\D\eta_1 \langle H_{\rm int,3}(\eta_1)H_{\rm int,3}(\eta_2) \pi_{k_1}(\eta_0)\pi_{k_2}(\eta_0)\pi_{k_3}(\eta_0)\pi_{k_4}(\eta_0)\rangle}\\&+ \int_{-\infty}^{\eta_0}d\D\eta_1 \int_{-\infty}^{\eta_0}\D\eta_2 \langle H_{\rm int,3}(\eta_1)\pi_{k_1}(\eta_0)\pi_{k_2}(\eta_0)\pi_{k_3}(\eta_0)\pi_{k_4}(\eta_0)H_{\rm int,3}(\eta_2)\rangle,
    \end{aligned}
\end{equation}
where the cubic Hamiltonian is given by
\begin{equation}
    \begin{aligned}
        H_{\rm int,3}&=\frac{\beta a}{2M^2} \int \D^3x \partial_\eta \pi (\partial \pi)^2\\&=-\frac{\beta a}{2M^2} \int \frac{\D^3\bm p_1}{(2\pi)^3} \int \frac{\D^3\bm p_2}{(2\pi)^3} \int \frac{\D^3\bm p_3}{(2\pi)^3} (2\pi)^3 \delta(\Sigma_{i=1}^3\bm p_i) (\bm p_2\cdot \bm p_3) \partial_\eta \pi_{\bm p_1} \pi_{\bm p_2} \pi_{\bm p_3}\,.
    \end{aligned}
\end{equation}
We calculate the two terms in \eqref{trispectrum_SE} separately. The first term is
\begin{equation}
    \begin{aligned}
        &\braket{\pi_{\bm k_1}(\eta_0)\pi_{\bm k_2}(\eta_0)\pi_{\bm k_3}(\eta)\pi_{\bm k_4}(\eta_0)}\\
        &=2\Re{-\int_{-\infty}^{\eta_0}d\eta _2\int_{-\infty}^{\eta_2}d\eta_1 \langle H_{\rm int,3}(\eta_1)H_{\rm int,3}(\eta_2) \pi_{k_1}(\eta_0)\pi_{k_2}(\eta_0)\pi_{k_3}(\eta_0)\pi_{k_4}(\eta_0)\rangle}.
    \end{aligned}
\end{equation}
We then get 
\begin{equation}
    \begin{aligned}
        &T_{\pi,1}(\bm k_1,\bm k_2,\bm k_3,\bm k_4)\\
        &=-\frac{\beta^2H^8} {M^4}\frac{\pi\Gamma(\frac34)^4}{2^{11}}(q_1q_2q_3q_4)^{-\frac34} \Re \int_{-\infty}^{0}\D\eta_2\int_{-\infty}^{\eta_2}\D\eta_1 (-\eta_1)^{\frac92}(-\eta_2)^{\frac92}\\&
        \left[(\bm k_1\cdot \bm k_2)(\bm k_3\cdot \bm k_4) q_{12}q_{34} H_{-\frac14}^{(1)}(q_{12}\eta_1^2)H_{\frac34}^{(1)}(q_1\eta_1^2)H_{\frac34}^{(1)}(q_2\eta_1^2) H_{-\frac14}^{(2)}(q_{34}\eta_2^2)H_{\frac34}^{(1)}(q_3\eta^2_2)H^{(1)}_{\frac34}(q_4\eta_2^2) \right.\\&+23 \text{perms}\\
        &-(\bm k_1\cdot \bm k_2)\bm k_4\cdot \bm k_{34}q_{12}q_3 H_{-\frac14}^{(1)}(q_{12}\eta_1^2)H_{\frac34}^{(1)}(q_1\eta_1^2)H_{\frac34}^{(1)}(q_2\eta_1^2) H_{-\frac14}^{(1)}(q_3\eta_2^2)H_{\frac34}^{(1)}(q_4\eta_2^2) H_{\frac34}^{(2)}(q_{34}\eta_2^2)\\&+47 \text{perms}\\
        &-\bm k_2\cdot\bm k_{12}\bm k_3\cdot \bm k_4 q_1q_{34} H_{-\frac14}^{(1)}(q_1\eta_1^2)H_{\frac34}^{(1)}(q_2\eta_1^2)H_{\frac34}^{(1)}(q_{12}\eta_1^2)H_{-\frac14}^{(2)}(q_{34}\eta_2^2)H_{\frac34}^{(1)}(q_3\eta_2^2)H_{\frac34}^{(1)}(q_4\eta_2^2)\\&+47 \text{perms}\\
        &\left. +\bm k_1\cdot\bm k_{12}\bm k_4\cdot\bm k_{34}H_{-\frac14}^{(1)}(q_1\eta_1^2)H_{\frac34}^{(1)}(q_2\eta_1^2)H_{\frac34}^{(1)}(q_{12}\eta_1^2)H_{-\frac14}^{(1)}(q_3\eta_2^2)H_{\frac34}^{(1)}(q_4\eta_2^2)H_{\frac34}^{(2)}(q_{34}\eta_2^2)\right.\\&+95 \text{perms}\Big],
    \end{aligned}
\end{equation}
where we defined $\bm k_{ij}\equiv \bm k_i+\bm k_j$ and $q_{ij}=\frac{\sqrt\alpha H}{2M}(k_i+k_j)^2$.
Besides, we have taken $\eta_0\to 0$, used \eqref{pi_solu} and
\begin{equation}
    \pi_k(\eta)|_{\eta\rightarrow 0}=-i\frac{\Gamma(\frac34)}{\sqrt \pi 2^{\frac34}} H q^{-\frac34}.
\end{equation}
Note that $H_{-1/4}^{(1)}$ comes from $\pi'$,
\begin{equation}
    \partial_\eta\pi_k(\eta)=-2qH\sqrt{\frac\pi8} (-\eta)^{\frac52} H_{-1/4}^{(1)}(q\eta^2).
\end{equation}
To proceed and perform the integral, we change variables to $x_1=(-\frac{\eta_1}{\eta_2})^2$ and $x_2=\frac{\sqrt{\alpha}H}{2M}k^2 \eta_2^2$, where $k$ is a positive constant with dimension of momentum.
After some algebra we arrive at
\begin{equation}
    \begin{aligned}
                T_{\pi,1}&=-\beta^2\alpha^{-\frac{13}{4}} H^4\left(\frac HM\right)^{-\frac52} \frac{\pi\Gamma(\frac34)^4}{2^6\sqrt2}
                \left({\hat k_1\hat k_2\hat k_3\hat k_4}\right)^{-3/2}k^{-9}\\
                &\Re \Bigg[ i^{11/2}\int_{0}^{\infty}\D x_2\int_{1}^{\infty}\D x_1 x_1^{\frac74}x_2^{\frac92} \left( {
                \hat{\bm k}_1\cdot \hat{\bm k}_2}{\hat{\bm k}_3\cdot \hat{\bm k}_4}\hat{ k}_{12}^2{\hat k_{34}^2} \right.\\
                & H_{-\frac14}^{(1)}(i\hat{k}_{12}^2x_1x_2) H_{\frac34}^{(1)}(i\hat k_1^2x_1x_2) H_{\frac34}^{(1)}(i\hat{k}_2^2x_1x_2) H_{-\frac14}^{(2)}(i\hat{k}_{34}^2x_2)H_{\frac34}^{(1)}(i\hat k_3^2x_2) H_{\frac34}^{(1)}(i\hat k_4^2x_2)+23 \text{perms}\\
                &-\hat{\bm k}_1\cdot \hat{\bm k}_2 \hat{\bm k}_4\cdot\hat{\bm k}_{34}\hat{k}_{12}^2\hat k_3^2\\
                & H_{-\frac14}^{(1)}(i\hat{k}_{12}^2x_1x_2) H_{\frac34}^{(1)}(i\hat{k}_1^2x_1x_2) H_{\frac34}^{(1)}(i\hat{k}_2^2x_1x_2)H_{-\frac14}^{(1)}(i\hat{k}_3^2x2) H_{\frac34}^{(1)}(i\hat{k}_4^2x_2) H_{\frac34}^{(2)}(i\hat{k}_{34}^2x_2)+47 \text{perms} \\& 
                -\hat{\bm k}_2\cdot\hat{\bm k}_{12}\hat{\bm k}_3\cdot\hat{\bm k}_4\hat{k}_1^2\hat{k}_{34}^2 \\
        & H_{-\frac14}^{(1)}(i\hat{k}_1^2x_1x_2)H_{\frac34}^{(1)}(i\hat{k}_2^2x_1x_2) H_{\frac34}^{(1)}(i\hat{k}_{12}^2x_1x_2) H_{-\frac14}^{(2)}(i\hat{k}_{34}^2x_2)H_{\frac34}^{(1)}(i\hat{k}_3^2x_2) H_{\frac34}^{(1)}(i\hat{k}_4^2x_2)+47 \text{perms} \\&
        \hat{\bm k}_2\cdot\hat{\bm k}_{12}\hat{\bm k}_4\cdot\hat{\bm k}_{34}\hat{k}_1^2\hat{k}_3^2
        \\
        & \left. H_{-\frac14}^{(1)}(i\hat{k}_1^2 x_1x_2) H_{\frac34}^{(1)}(i\hat{k}_2^2x_1x_2) H_{\frac34}^{(1)}(i\hat{k}_{12}^2x_1x_2) H_{-\frac{1}{4}}^{(1)}(i\hat{k}_3^2x_2)H_{\frac34}^{(1)}(i\hat{k}_4^2x_2)H_{\frac34}^{(2)}(i\hat{k}_{34}^2x_2)+95\text{perms} \right)
                \Bigg],
    \end{aligned}
\end{equation}
where we denoted $\hat{\bm k}_i\equiv \bm k_i/k$, and Wick rotated $x_2\to ix_2$ as there is no pole in the first quadrant.
Now we can use the following relations to extract the real part:
\begin{equation}\label{Hankel_to_Bessel}
    H_\nu^{(1)}(iz)=\frac{2}{i\pi}e^{-i\nu\pi/2}K_\nu(z),\quad H_{\nu}^{(2)}(iz)=\frac{ie^{-i\frac{\nu\pi}{2}}}{\sin{\nu\pi}}\left( (1-e^{2i\nu\pi}) I_{\nu}(z)+\frac{2}{\pi}\sin{\nu\pi}K_{\nu}(z)\right).
\end{equation}
Taking into account $\zeta=-H\pi/M^2$, we finally get
\begin{equation}
     \begin{aligned}
        T_{\zeta,1}=& -\beta^2 \alpha^{-\frac{13}{4}} \left(\frac HM\right)^{\frac{11}{2} }\frac{\Gamma(\frac34)^4}{2\pi^4} \left({\hat k_1\hat k_2\hat k_3\hat k_4}\right)^{-3/2} k^{-9} \\
        & \int_{0}^{\infty} \D x_2\int_{1}^{\infty}\D x_1 x_1^{\frac74}x_2^{\frac92} \left(-
        {\hat{\bm k}_1\cdot \hat{\bm k}_2}{\hat{\bm k}_3\cdot \hat{\bm k}_4}\hat{ k}_{12}^2{\hat k_{34}^2} \right.\\
        & K_{\frac14}(\hat{k}_{12}^2 x_1x_2) K_{\frac34} (\hat{k}_{1}^2x_1x_2) K_{\frac34} (\hat{k}_{2}^2x_1x_2) I_{-\frac14}(\hat{k}_{34}^2 x_2) K_{\frac34}(\hat{k}_{3}^2x_2)K_\frac{3}{4}(\hat{k}_{4}^2x_2) +23 \text{perms}\\& 
-\hat{\bm k}_1\cdot \hat{\bm k}_2 \hat{\bm k}_4\cdot\hat{\bm k}_{34}\hat{k}_{12}^2\hat k_3^2\\
& K_{\frac14}(\hat{k}_{12}^2x_1x_2) K_{\frac34}(\hat{k}_{1}^2x_1x_2)K_{\frac34}(\hat{k}_{2}^2x_1x_2) K_{\frac14}(\hat{k}_{3}^2x_2)K_{\frac34}(\hat{k}_{4}^2x_2) I_{\frac34}(\hat{k}_{34}^2x_2)+47 \text{perms}\\
&+\hat{\bm k}_2\cdot\hat{\bm k}_{12}\hat{\bm k}_3\cdot\hat{\bm k}_4\hat{k}_1^2\hat{k}_{34}^2
\\
& K_{\frac14}(\hat{k}_{1}^2x_1x_2)K_{\frac34}(\hat{k}_{2}^2x_1x_2)K_{\frac34}(\hat{k}_{12}^2x_1x_2) I_{-\frac14}(\hat{k}_{34}^2x_2) K_{\frac{3}{4}}(\hat{k}_{3}^2x_2)K_{\frac34}(\hat{k}_{4}^2x_2)+47 \text{perms}\\
&+\hat{\bm k}_2\cdot\hat{\bm k}_{12}\hat{\bm k}_4\cdot\hat{\bm k}_{34}\hat{k}_1^2\hat{k}_3^2
 \\
& K_{\frac14}(\hat{k}_{1}^2x_1x_2)K_{\frac34}(\hat{k}_{2}^2x_1x_2)K_{\frac34}(\hat{k}_{12}^2x_1x_2)K_{\frac14}(\hat{k}_{3}^2x_2)K_{\frac34}(\hat{k}_{4}^2x_2)I_{\frac34}(\hat{k}_{34}^2x_2)+95 \text{perms}\Big).
     \end{aligned}
 \end{equation}

Now we turn to the second term in \eqref{trispectrum_SE}. 
With similar procedures as above we find
 \begin{equation}
     \begin{aligned}
         T_{\pi,2}=&\beta^2\alpha^{-\frac{13}{4}} H^4\left(\frac HM\right)^{-\frac52} \frac{\pi\Gamma(\frac34)^4}{2^7\sqrt2} 
         \left({\hat k_1\hat k_2\hat k_3\hat k_4}\right)^{-3/2} k^{-9} \\\Bigg[
         & \hat{\bm k}_1\cdot \hat{\bm k}_2\hat{k}_{12}^2 \int_{0}^\infty \D x_{1} x_1^{\frac74} H_{-\frac14}^{(1)}(\hat{k}_{12}^2 x_1) H_{\frac34}^{(1)}(\hat{k}_{1}^2 x_1) H_{\frac34}^{(1)}(\hat{k}_{2}^2 x_1)\\
         &\hat{\bm k}_3\cdot \hat{\bm k}_4 \hat{k}_{34}^2 
         \int_{0}^{\infty}dx_2 x_2^{\frac74} H_{-\frac14}^{(2)}(\hat{k}_{34}^2x_2) H_{-\frac14}^{(2)}(\hat{k}_{3}^2x_2) H_{\frac34}^{(2)}(\hat{k}_{4}^2x_2) +23 \text{perms} \\
         & - \hat{\bm k}_1\cdot\hat{\bm k}_2\hat{k}_{12}^2
         \int_{0}^\infty \D x_{1} x_1^{\frac74} H_{-\frac14}^{(1)}(\hat{k}_{12}^2 x_1) H_{\frac34}^{(1)}(\hat{k}_{1}^2 x_1) H_{\frac34}^{(1)}(\hat{k}_{2}^2 x_1)\\
         & \hat{\bm k}_4\cdot\hat{\bm k}_{34}\hat{k}_{3}^2
         \int_{0}^{\infty}\D x_2 (x_2)^{\frac74} H_{-\frac14}^{(2)}(\hat{k}_{3}^2x_2)H_{\frac34}^{(2)}(\hat{k}_{4}^2 x_2)H_{\frac34}^{(2)}(\hat{k}_{34}^2x_2)+95 \text{perms} \\
         & +\hat{\bm k}_2\cdot\hat{\bm k}_{12}\hat{k}_1^2
         \int_{0}^{\infty}\D x_1 x_1^{\frac74}H_{-\frac14}^{(1)}(\hat{k}_{1}^2x_1)H_{\frac34}^{(1)}(\hat{k}_{2}^2x_1) H_{\frac34}^{(1)}(\hat{k}_{12}^2x_1) \\
         & \hat{\bm k}_4\cdot\hat{\bm k}_{34}\hat{k}_3^2
          \int_{0}^{\infty}\D x_2 (x_2)^{\frac74} H_{-\frac14}^{(2)}(\hat{k}_{3}^2x_2)H_{\frac34}^{(2)}(\hat{k}_{4}^2 x_2)H_{\frac34}^{(2)}(\hat{k}_{34}^2x_2)+95 \text{perms} \Bigg].
     \end{aligned}
 \end{equation}
 Now we perform Wick rotations, $x_1\to ix_1$ and $x_2\to -ix_2$. Note that we rotate $x_2$ into the negative imaginary axis as $H_\nu ^{(2)}$ is identically vanishing at the infinity of the fourth quadrant. We then obtain the following trispectrum for $\zeta$:
  \begin{equation}
     \begin{aligned}
         T_{\zeta,2}&=\beta^2\alpha^{-\frac{13}{4}}\left(\frac HM\right)^{\frac{11}{2}} \frac{\Gamma(\frac34)^4}{2\sqrt2\pi^5} \left({\hat k_1\hat k_2\hat k_3\hat k_4}\right)^{-\frac32} k^{-9} \\
         &  \hat{\bm k}_1\cdot \hat{\bm k}_2\hat{k}_{12}^2
          \int_{0}^\infty\D x_{1} x_1^{\frac74} K_{\frac14}(\hat{k}_{12}^2x_1) K_{\frac34}(\hat{k}_{1}^2x_1) K_{\frac34}(\hat{k}_{2}^2x_1) \\
         & \hat{\bm k}_3\cdot\hat{\bm k}_4\hat{k}_{34}^2
         \int_{0}^{\infty}\D x_2 x_2^{\frac74} K_{\frac14}(\hat{k}_{34}^2x_2) K_{\frac34}(\hat{k}_{3}^2x_2) K_{\frac34}(\hat{k}_{4}^2x_2)+23 \text{perms} \\
         &  - \hat{\bm k}_1\cdot \hat{\bm k}_2\hat{k}_{12}^2
         \int_{0}^\infty \D x_{1} x_1^{\frac74} K_{\frac14}(\hat{k}_{12}^2x_1) K_{\frac34}(\hat{k}_{1}^2x_1) K_{\frac34}(\hat{k}_{2}^2x_1)\\
         &\hat{\bm k}_3\cdot\hat{\bm k}_{34}\hat{k}_3^2
         \int_{0}^{\infty}\D x_2 x_2^{\frac74} K_{\frac14}(\hat{k}_{3}^2x_2) K_{\frac34}(\hat{k}_{4}^2x_2) K_{\frac34}(\hat{k}_{34}^2x_2)+95 \text{perms}\\
         & +\hat{\bm k_2}\cdot \hat{\bm k}_{12}\hat{k}_1^2
         \int_{0}^{\infty}\D x_1 x_1^{\frac74} K_{\frac14}(\hat{k}_{1}^2x_1) K_{\frac34}(\hat{k}_{2}^2x_1) K_{\frac34}(\hat{k}_{12}^2x_1)\\
         &\hat{\bm k}_3\cdot\hat{\bm k}_{34}\hat{k}_3^2
         \int_{0}^{\infty}\D x_2 x_2^{\frac74} K_{\frac14}(\hat{k}_{3}^2x_2) K_{\frac34}(\hat{k}_{4}^2x_2) K_{\frac34}(\hat{k}_{34}^2x_2)+95 \text{perms}.
     \end{aligned}
 \end{equation}

\subsection{Parity-even contact term}
The contact term contribution to the 4-point function can be written as
\begin{equation}\label{PE_contact-inin}
\begin{aligned}
    \langle \pi_{\bm k_1}\pi_{\bm k_2}\pi_{\bm k_3}\pi_{\bm k_4}\rangle =\frac{\gamma}{4M^4}\Re{i\int_{-\infty}^{\eta_0}\D\eta \int \D^3x \langle  (\partial_i\pi\partial_i\pi)^2 \pi_{\bm k_1}\pi_{\bm k_2}\pi_{\bm k_3}\pi_{\bm k_4}\rangle }.
\end{aligned}
\end{equation}
After some algebra we have
\begin{equation}
\begin{aligned}
    T_\pi(k_1,k_2,k_3,k_4)=& \frac{ \gamma H^8}{M^4} \frac{\Gamma(\frac34)^4}{2^8}(q_1q_2q_3q_4)^{-\frac34}\\&[(\bm k_1\cdot \bm k_2)(\bm k_3\cdot \bm k_4)+(\bm k_1\cdot\bm  k_3)(\bm k_2\cdot\bm  k_4)+(\bm k_1\cdot \bm k_4)(\bm k_2\cdot \bm k_3)]\\& \Re{i\int_{-\infty}^{0}\D \eta (-\eta)^6 H^{(1)}_{\frac34}(q_1\eta^2) H^{(1)}_{\frac34}(q_2\eta^2) H^{(1)}_{\frac34}(q_3\eta^2) H^{(1)}_{\frac34}(q_4\eta^2) }\\
    =& \gamma \alpha^{-\frac{13}{4}} H^4 \left( \frac HM \right) ^{-\frac52} \frac{\Gamma(\frac34)^4}{4\sqrt 2}(\hat{k}_1\hat k_2\hat k_3\hat k_4)^{-3/2}k^{-9}\\
    &\left[(\hat{\bm k}_1\cdot\hat{\bm k}_2)(\hat{\bm k}_3\cdot \hat{\bm k}_4)+(\hat{\bm k}_1\cdot \hat{\bm k}_3)(\hat{\bm k}_2\cdot \hat{\bm k}_4)+(\hat{\bm k}_1\cdot\hat{\bm k}_4)(\hat{\bm k}_2\cdot\hat{\bm k}_3)
    \right]\\
    &\Re{i\int_{0}^{\infty} \D z z^{5/2}H_{\frac34}^{(1)}(\hat{k}_1^2z) H_{\frac34}^{(1)}(\hat{k}_2^2z) H_{\frac34}^{(1)}(\hat{k}_3^2z) H_{\frac34}^{(1)}(\hat{k}_4^2z)},
\end{aligned}
\end{equation}
where we have changed variable as $z=\frac{\sqrt{\alpha} H}{2M}k^2\eta^2$.
Now we perform a Wick rotation $z\to iz$, and use \eqref{Hankel_to_Bessel} to get the final expression
\begin{equation}
\begin{aligned}
T_\zeta=& -  \gamma \alpha^{-\frac{13}{4}} \left( \frac HM \right) ^{\frac{11}{2}} \frac{2\Gamma(\frac34)^4}{\pi^4}(\hat{k}_1\hat k_2\hat k_3\hat k_4)^{-3/2}k^{-9}\\
&\left[(\hat{\bm k}_1\cdot\hat{\bm k}_2)(\hat{\bm k}_3\cdot \hat{\bm k}_4)+(\hat{\bm k}_1\cdot \hat{\bm k}_3)(\hat{\bm k}_2\cdot \hat{\bm k}_4)+(\hat{\bm k}_1\cdot\hat{\bm k}_4)(\hat{\bm k}_2\cdot\hat{\bm k}_3)
    \right]\\
& \int_{0}^{\infty} \D z z^{\frac52} K_{\frac34} (\hat{k}_1^2z) K_{\frac34}(\hat{k}_2^2z) K_{\frac34} (\hat{k}_3^2z)  k_{\frac34}(\hat{k}_4^2z).
\end{aligned}
\end{equation}
Through the equation \eqref{Ph_connected}, the tensor power spectrum is
\begin{equation}
\begin{aligned}
\mathcal P_{h,c}=&- \gamma\alpha^{-\frac{13}{4}} \left(\frac HM \right)^{\frac{11}{2}} \left(\frac{\mathcal H}{k} \right)^2 \frac{\Gamma(\frac34)^4}{2^5\pi^{11}} \\
&\int_{0}^{\infty} \D v_1 \int_{|1-v_1|}^{1+v_1} \D u_1 \int_{0}^{\infty} \D v_2 \int_{|1-v_2|}^{1+v_2} \D u_2 \int_{0}^{2\pi} \D \psi \mathcal{K}_c(u_1,v_1,u_2,v_2) \\
& 
\left[(\hat{\bm k}_1\cdot\hat{\bm k}_2)(\hat{\bm k}_3\cdot \hat{\bm k}_4)+(\hat{\bm k}_1\cdot \hat{\bm k}_3)(\hat{\bm k}_2\cdot \hat{\bm k}_4)+(\hat{\bm k}_1\cdot\hat{\bm k}_4)(\hat{\bm k}_2\cdot\hat{\bm k}_3)
    \right]\\
& \cos(2\psi) (u_1v_1u_2v_2)^{\frac34} \int_{0}^{\infty} \D z z^{\frac52} K_{\frac34} (u_1^2z) K_{\frac34}(v_1^2z) K_{\frac34} (u_2^2z)  k_{\frac34}(v_2^2z),
\end{aligned}
\end{equation}
in which the constant $k$ is identified as the external momentum, and $\hat k_{1,3}=v_{1,2}$ $\hat k_{2,4}=u_{1,2}$.

\subsection{First parity-odd contact term}

The 4-point function from the interaction \eqref{eq:PO lagrangians1} reads
\begin{equation}\label{PO_contact_inin}
    \begin{aligned}
       \langle \pi_{\bm k_1}\pi_{\bm k_2}\pi_{\bm k_3}\pi_{\bm k_4}\rangle&=2i\Im{ \langle i\int_{-\infty}^{\eta_0}\D \eta H_{\rm PO,1} \pi_{k_1}(\eta_0) \pi_{k_2}(\eta_0) \pi_{k_3}(\eta_0) \pi_{k_4}(\eta_0) \rangle},
    \end{aligned}
\end{equation}
where the Hamiltonian is 
\begin{equation}
    \begin{aligned}
        H_{\rm PO,1}&=-\frac{\gamma_1}{8M^{10}} \int \D^3x a^{-6}(\eta) \partial_\eta \pi(x,\eta) \epsilon_{ijk} \partial_{il} \pi(x,\eta) \partial_{lj}\partial^2 \pi(x,\eta) \partial_k \partial^2 \pi(x,\eta).
    \end{aligned}
\end{equation}
Note that since $H_{\rm PO,1}\to -H_{\rm PO,1}$ under change of signs of internal momenta $\bm p_i\to -\bm p_i$, so we have \eqref{PO_contact_inin} instead of \eqref{PE_contact-inin}. 
This is a feature of parity-odd interactions.

With the same procedures as before, we get the trispectrum for $\zeta$,
\begin{equation}
    \begin{aligned}
        &T_\zeta(\bm k_1,\bm k_2,\bm k_3,\bm k_4)\\
         &= -i\gamma_1\alpha^{-\frac92} \left(\frac{H}{M}\right)^9 \frac{4\Gamma(\frac34)^4}{\pi^4} k_1^{\frac12} k_2^{-\frac32} k_3^{\frac12} k_4^{\frac12} 
        (\bm k_2\cdot\bm k_3) \bm k_2\cdot (\bm k_3\times \bm k_4) k^{-14}\\& 
        \int _{0}^{\infty} \D z z^6 K_{\frac14}(\hat{k}_1^2z) K_{\frac34}(\hat{k}_2^2z) K_{\frac34}(\hat{k}_3^2z) K_{\frac34}(\hat{k}_4^2z) +23 \text{perms}.
    \end{aligned}
\end{equation}
And the difference between the power speactra of right-handed and left-handed polarizations is given by
\begin{equation}
    \begin{aligned}
        \mathcal P_R&-\mathcal P_L\\
        =& \gamma_1\alpha^{-\frac92} \left(\frac HM \right)^9 \left(\frac {\mathcal{H}} {k} \right)^2 \frac{4\Gamma(\frac34)^4}{\pi^5} \frac{1}{(2\pi)^6}  \int_{0}^{\infty} \D v_1 \int_{|v_1-1|}^{1+v_1} \D u_1 \int_{0}^{\infty} \D v_2  \\
        &\int_{|v_2-1|}^{1+v_2} \D u_2 \int_{0}^{2\pi} \D\psi \int_{0}^{\infty} \D z \mathcal K_c(u_1,v_1,u_2,v_2) \sin(2\psi) (v_1u_1v_2u_2)^{11/4} \\
        &  \left[ u_1^{-2}
        (\hat{\bm k}_2\cdot\hat{\bm k}_3)(\hat{\bm k}_3\cdot\hat{\bm k}_3\times \hat{\bm k}_4)z^6 K_{\frac14}(v_1^2z) K_{\frac34}(u_1^2z) K_{\frac34}(u_2^2z) K_{\frac34}(v_2^2z)
         +23\text{perms}  \right].
    \end{aligned}
\end{equation}

\subsection{Second parity-odd contact term}

Now we consider the trispectrum from the second parity-odd quartic interaction in \eqref{eq:PO lagrangians2},
\begin{equation}
    \langle \pi_{\bm k_1}\pi_{\bm k_2}\pi_{\bm k_3}\pi_{\bm k_4}\rangle=2i\Im{\langle i\int _{-\infty}^{\eta_0} \D\eta H_{\rm PO,2}\pi_{k_1}\pi_{k_2}\pi_{k_3}\pi_{_4}\rangle},
\end{equation}
where the interaction Hamiltonian is given by
\begin{equation}
    H_{\rm PO,2}=-\frac{\gamma_2}{8M^9} \int \D^3x a^{-5}(\eta) \epsilon_{ijk} \partial_{mn} \pi(x,\eta) \partial_{ni}\pi(x,\eta) \partial _{mlj} \pi \partial_{lk} \pi.
\end{equation}
With some algebra, change of variables, translating Hankel functions into Bessel functions and Wick rotation, we arrive at
\begin{equation}
    \begin{aligned}
        T_\zeta&= -i \gamma_2\alpha^{-\frac92} \left( \frac HM \right)^8 \frac{2\Gamma(\frac34)^4}{\pi^4} (\hat{k}_1\hat{k}_2\hat{k}_3\hat{k}_4)^{-\frac32}k^{-9} \\
        &\left( (\hat{\bm k}_1\cdot\hat{\bm k}_2)(\hat{\bm k}_1\cdot\hat{\bm k}_3)(\hat{\bm k_3}\cdot\hat{\bm k}_4)(\hat{\bm k_2}\cdot\hat{\bm k}_3\times \hat{\bm k}_4)
        +23 \text{perms} \right) \\
        & \int _{0}^{\infty} \D z z^5 K_{\frac34}(\hat{k}_1^2z) K_{\frac34}(\hat{k}_2^2z) K_{\frac34}(\hat{k}_3^2z) K_{\frac34}(\hat{k}_4^2z),
    \end{aligned}
\end{equation}
and
\begin{equation}
    \begin{aligned}
        &\mathcal P_R-\mathcal P_L=\\&\gamma_2\alpha^{-\frac92} \left( \frac HM \right)^8 \left(\frac{\mathcal H}{k} \right)^2 \frac{2\Gamma(\frac34)^4}{\pi^5} \frac{1}{(2\pi)^6} \int_{0}^{\infty}\D v_1 \int_{|v_1-1|}^{1+v_1}\D u_1 \int_{0}^{\infty}\D v_2 \int_{|v_2-1|}^{1+v_2}\D u_2 \int_{0}^{2\pi} \D \psi \int_{0}^{\infty} \D z \\
        &
        \left( 
        (\hat{\bm k}_1\cdot \hat{\bm k}_2)(\hat{\bm k}_1\cdot\hat{\bm k}_3)(\hat{\bm k}_3\cdot\hat{\bm k}_4)(\hat{\bm k}_2\cdot\hat{\bm k}_3\times\hat{\bm k}_4)
        +23 \text{perms} 
        \right) \\
        & \mathcal K_c(u_1,v_1,u_2,v_2) \sin(2\psi) (u_1v_1u_2v_2)^{\frac34} z^5 K_{\frac34}(u_1^2z) K_{\frac34}(v_1^2z) K_{\frac34}(u_2^2z) K_{\frac34}(v_2^2z),
    \end{aligned}
\end{equation}
where $\hat k_{1,3}=v_{1,2}$ $\hat k_{2,4}=u_{1,2}$.

\section{Notes on the numerics} \label{sec:app numerics}

We calculate integrals numerically with the Monte Carlo code Vegas+ \cite{Lepage:1977sw,Lepage:2020tgj}. Each integral is computed using 20 iterations of the algorithm and with $5\times 10^6$ (for the scalar exchange diagram) or $10^6$ (for the contact diagrams) points in the integration domain for each iteration.

In addition to the need to recast the integrals in terms of manifestly real functions and variables, as detailed in the previous Appendix, it is also useful to switch variables so as to integrate over a rectangular domain. To this end we use
\begin{equation}
u_i=\frac{t_i+s_i+1}{2} \,,\qquad v_i=\frac{t_i-s_i+1}{2} \,,
\end{equation}
which gives
\begin{equation}
\int_0^{\infty}\D v_i\int_{|1-v_i|}^{1+v_i}\D u_i(\ldots)=\frac{1}{2}\int_0^{\infty}\D t_i\int_{-1}^1\D s_i(\ldots) \,.
\end{equation}

The following are our results, given in truncated form in the main text, for the dimensionless master integrals for the four trispectra considered in this work:
\begin{equation}
\begin{aligned}
\mathcal{I}_{\rm PE1}&=-0.0024221(43)\,,\\
\mathcal{I}_{\rm PE2}&=-0.0203791(75) \,,\\
\mathcal{I}_{\rm PO1}&=-0.33466(92)\,, \\
\mathcal{I}_{\rm PO2}&=-0.81655(24) \,.
\end{aligned}
\end{equation}
For each outcome, the $N$ digits in parenthesis denote the error in the last $N$ digits of the result. Our numerical estimates therefore have a relative precision of order $10^{-4}-10^{-3}$.

%==========================

% \bibliographystyle{JHEP}
% \bibliography{ref}

\providecommand{\href}[2]{#2}\begingroup\raggedright\endgroup

\end{document}